\documentclass[a4paper,11pt]{article}
\pdfoutput=1 
\usepackage{jheppub} 
\usepackage[T1]{fontenc} 
\usepackage{amssymb}
\usepackage{graphicx}
\usepackage{color}
\usepackage{tabls}
\usepackage{hyperref}
\usepackage{bm}
\usepackage{orcidlink}
\newcommand{\Z}{\mathbb{Z}}
\newcommand{\U}{\mathrm{U}}
\newcommand{\SU}{\mathrm{SU}}
\newcommand{\redchisq}{\chi^2_{\tiny\mbox{red}}}
\newcommand{\betac}{\beta_{\mbox{\tiny{c}}}}
\newcommand{\Ntau}{N_\tau}
\newcommand{\Tc}{T_{\mbox{\tiny{c}}}}
\newcommand{\mg}{m_{\mbox{\tiny{g}}}}
\newcommand{\Ng}{N_{\mbox{\tiny{g}}}}
\newcommand{\eq}[1]{\begin{align}\label{#1}}
\newcommand{\en}{\end{align}}
\newcommand{\eqar}[1]{\begin{align}\label{#1}}
\newcommand{\enar}{\end{align}}

\usepackage[english]{babel}
\usepackage{lmodern}

\usepackage{xcolor}

\usepackage{dsfont}
\usepackage{amsmath}
\usepackage{amsthm}
\usepackage{mathtools}
\usepackage{mathrsfs}
\usepackage{slashed}
\usepackage{physics}
\usepackage{tikz}
\usetikzlibrary{decorations.pathreplacing, decorations.markings, snakes}
\usetikzlibrary{arrows.meta}

\usepackage{afterpage}

\usepackage{caption}
\usepackage{subcaption}
\usepackage{float}
\usepackage[format=plain,
            font=it]{caption}
\usepackage{multirow}

\newcommand{\lam}{|\Lambda|}

\title{\boldmath Duality transformations and the entanglement entropy of gauge theories}

\author{Andrea~Bulgarelli\orcidlink{0009-0002-2917-6125}}
\author{and Marco~Panero\orcidlink{0000-0001-9477-3749}}

\affiliation{Physics Department, University of Turin \& INFN, Turin unit\\Via Pietro Giuria 1, I-10125 Turin, Italy}

\emailAdd{andrea.bulgarelli@unito.it}
\emailAdd{marco.panero@unito.it}

\abstract{The study of entanglement in gauge theories is expected to provide insights into many fundamental phenomena, including confinement. However, calculations of quantities related to entanglement in gauge theories are limited by ambiguities that stem from the non-factorizability of the Hilbert space. In this work we study lattice gauge theories that admit a dual description in terms of spin models, for which the replica trick and R\'enyi entropies are well defined. In the first part of this work, we explicitly perform the duality transformation in a replica geometry, deriving the structure of a replica space for a gauge theory. Then, in the second part, we calculate, by means of Monte Carlo simulations, the entropic c-function of the $\Z_2$ gauge theory in three spacetime dimensions, exploiting its dual description in terms of the three-dimensional Ising model.}

\begin{document} 


\maketitle
\flushbottom

\section{Introduction}
\label{sec:introduction}

Entanglement is a key feature of quantum systems~\cite{Vidal:2002rm, Calabrese:2004eu, Kitaev:2005dm, Amico:2007ag} and has a very broad range of implications, ranging from those in condensed-matter physics~\cite{Osborne:2002zz, Latorre:2003kg, Laflorencie:2015eck} to those relevant for high-energy theory, black-hole physics, and holography~\cite{Sorkin:1984kjy, Bombelli:1986rw, Srednicki:1993im, Holzhey:1994we, Ryu:2006bv, Ryu:2006ef, Emparan:2006ni, Fursaev:2006ih, Myers:2010tj, VanRaamsdonk:2010pw, Liu:2012eea, Myers:2012ed, Maldacena:2013xja, Faulkner:2013yia, Faulkner:2013ana, Susskind:2014rva, Swingle:2014uza, Harlow:2014yka, Rangamani:2016dms}. Finally, entanglement is also an important quantum resource in quantum information theory~\cite{nielsen2000quantum, Dittel:2023nso}: for example, entanglement distillation can be used in quantum error-correction~\cite{Bennett:1995ra, Bennett:1995tk}.

In general, defining entanglement measures for multipartite and mixed states involves some difficulties~\cite{Horodecki:2009zz}, but in the case of bipartite entanglement for pure states there exist well established measures, among which the entanglement entropy is one of the most studied. Entanglement entropy can be thought of as the amount of entropy associated with the degrees of freedom in a spatial region $A$ of a quantum system, and is usually defined in terms of the von~Neumann entropy $S = -\Tr(\rho_A\ln\rho_A)$ of the reduced density matrix $\rho_A$, obtained by tracing the density matrix $\rho$ of the system over the degrees of freedom of the complement of $A$ (that we denote as $B$): $\rho_A = \Tr_B\rho$. It should be noted that this definition involves non-trivial subtleties~\cite{Witten:2018zxz}: in particular, it relies on the hypothesis that the Hilbert space of the theory factorizes into a direct product of the Hilbert spaces defined for the subsystem $A$ and for its complement, $\mathcal{H} = \mathcal{H}_A\otimes\mathcal{H}_B$, which, strictly speaking, is not satisfied for a quantum field theory in continuum, nor for gauge theories---not even when they are regularized on a lattice. When the factorization of the Hilbert space holds, the entanglement entropy can be expressed as the $n\rightarrow 1$ limit of the R\'enyi entropy of order $n$ associated to the subsystem $A$:
\begin{align}
\label{Renyi_entropy_definition}
S_n = \frac{1}{1-n}\ln\Tr\rho^n_A.
\end{align} 
In quantum field theories the entanglement entropy is affected by ultraviolet divergences: for highly excited states the divergence is proportional to the spatial volume of the system~\cite{Bianchi:2021aui}, whereas ground states exhibit the so-called ``area law''~\cite{Eisert:2008ur}, namely the ultraviolet divergence is proportional to the area of the entangling surface $\partial A$ separating $A$ and $B$. Therefore, a proper regularization is required, in order to identify the subleading contributions, which encode important physical information about the ``resolution scale'' at which the degrees of freedom of the system are probed~\cite{Casini:2010kt, Liu:2012eea, Casini:2012ei, Myers:2012ed, Klebanov:2012va, Liu:2013una, Jokela:2024cxb}, the geometry of the system~\cite{Klebanov:2012yf, Casini:2014yca, Mezei:2014zla, Faulkner:2015csl, Bianchi:2016xvf} and the nature of the state itself~\cite{Casini:2008cr}. For a system defined in $D=d+1$ spacetime dimensions, it is then convenient to consider a geometry in which the entangling surface does not depend on the size of the subsystem $A$; a common choice is the slab geometry, in which $A$ is maximally extended in all space-like directions except for the one separating $A$ and $B$, where it has a finite size $l$. One can then define the ultraviolet-finite entropic (R\'enyi) c-function as~\cite{Nishioka:2006gr, Casini:2006es, Casini:2004bw}
\begin{align}
C_n(l) = \frac{l^{D-1}}{|\partial A|}\frac{\partial S_n}{\partial l},
\label{definition_entropic_c-function_slab}
\end{align}
where $|\partial A|$ denotes the area of the entangling surface. It has been proven in a large number of cases that the entropic c-function is monotonically decreasing along the renormalization group flow~\cite{Casini:2004bw, Casini:2012ei, Casini:2017vbe, Casini:2023kyj}, providing a measure of the effective number of degrees of freedom of a theory~\cite{Zamolodchikov:1986gt}.

Calculations of such highly non-local quantities are typically challenging, both analytically and numerically. A tool which is commonly used is the replica trick~\cite{Calabrese:2004eu,Calabrese:2009qy}, that allows one to express the trace of a power of the reduced density matrix appearing in eq.~\eqref{Renyi_entropy_definition} in terms of ratios of partition functions,
\begin{align}
\Tr\rho_A^n = \frac{Z_n}{Z^n},
\end{align}
where $Z^n$ denotes the product of the partition functions of $n$ independent copies of the system, while $Z_n$ is the partition function of the theory defined on a Riemann surface, obtained connecting different replicas together through a cut corresponding to the subsystem $A$ at fixed time. The manifold thus defined has a conical singularity corresponding to the entangling surface $\partial A$. This method, originally designed for analytical calculations, has also been implemented in lattice simulations~\cite{Buividovich:2008kq, Buividovich:2008gq, Buividovich:2008yv, Caraglio:2008pk, Alba:2009ek, Gliozzi:2009zc, Nakagawa:2009jk, Nakagawa:2010kjk, Hastings:2010zka, Alba:2011fu, Humeniuk:2012xg, Grover:2013nva, Coser:2013qda, Drut:2015aoa, Itou:2015cyu, Alba:2016bcp, Rabenstein:2018bri, DEmidio:2019usm, Rindlisbacher:2022bhe, Bringewatt:2023xxc, Jokela:2023rba, Amorosso:2023fzt, Bulgarelli:2023ofi, Bulgarelli:2023fgv}. On a Euclidean lattice, the replica geometry is directly implemented by means of different boundary conditions along the Euclidean-time direction for the subsystem $A$ and for its complement $B$, namely the period of $A$ is $n$ times the period of $B$.

Another approach that is widely used in literature to study the entanglement entropy of theories that admit a holographic dual~\cite{Maldacena:1997re, Gubser:1998bc, Witten:1998qj} is to use the Ryu--Takayanagi formula~\cite{Ryu:2006bv, Ryu:2006ef, Emparan:2006ni, Fursaev:2006ih, Nishioka:2006gr}, which relates the entanglement entropy of a theory defined on the boundary to minimal-area surfaces extending in the bulk. This approach led to important predictions for the behavior of entanglement in strongly coupled gauge theories. In particular, in ref.~\cite{Klebanov:2007ws} it was conjectured that entanglement can be a probe to study the phenomenon of confinement (see also refs.~\cite{Fujita:2008zv, Lewkowycz:2012mw, Kol:2014nqa, Jokela:2020wgs}); this possibility has been investigated numerically, by means of Monte Carlo lattice simulations, in $\SU(N)$ gauge theories~\cite{Buividovich:2008kq, Buividovich:2008gq, Buividovich:2008yv, Nakagawa:2009jk, Nakagawa:2010kjk, Itou:2015cyu, Rabenstein:2018bri}.

A non-trivial aspect in the study of entanglement entropy in gauge theories, however, is the ambiguity of the definition itself in the presence of a local symmetry. Gau{\ss}'s law and consequently the non-local constraint defining physical states makes the physical Hilbert space of a gauge theory non-factorizable. This problem has been extensively discussed in the literature, and different definitions of entanglement entropy have been proposed~\cite{Buividovich:2008gq, Donnelly:2011hn, Agon:2013iva, Casini:2013rba, Radicevic:2014kqa, Ohmori:2014eia, Donnelly:2014gva, Ghosh:2015iwa, Aoki:2015bsa, Chen:2015kfa, Radicevic:2015sza, Soni:2015yga, VanAcoleyen:2015ccp, Radicevic:2016tlt, Soni:2016ogt, Aoki:2016lma, Lin:2018bud}, among which two approaches are most commonly used: one can either extend the Hilbert space of states, including gauge non-invariant ones to ensure factorizability; or work at the level of the algebra of observables defined over a specific region, specifying which generators have to be included in region $A$ and which ones in its complement. It turns out that for some choices of the algebras the two approaches are equivalent~\cite{Soni:2015yga}, and lead to the following general result: the trace which defines the entanglement entropy splits into the sum of different contributions, labeled by the eigenvalues of the operators at the boundary between the system $A$ and its complement, which in general belong to the center of the algebra of observables of $A$. As a result, in Abelian gauge theories defined on the lattice the entanglement entropy admits the following decomposition
\begin{align}
S = -\sum_k p^{(k)}\ln p^{(k)} - \sum_k p^{(k)} S(\rho^{(k)}_A),
\label{general_expression_entanglement_entropy_gauge_theories}
\end{align}
where $k$ labels different sectors, determined by the eigenvalues of the boundary operators, $p^{(k)}$ denotes the probability for the sector $k$, while $S(\rho^{(k)}_A)$ is the Shannon entropy of the reduced density matrix restricted to the sector $k$. For theories based on a non-Abelian gauge group, eq.~\eqref{general_expression_entanglement_entropy_gauge_theories} includes a further term, which is related to the dimension of the representation of the electric flux at the boundary.

From eq.~\eqref{general_expression_entanglement_entropy_gauge_theories} one might worry that the entropy thus defined is not really quantifying the entanglement between two subsystems, since only the second term is distillable, while the first one is related to the classical probability distribution of the eigenvalues of the boundary operators, which depends on the specific (and somehow arbitrary) choice of operators that are included in $A$. While this ambiguity is indeed present on the lattice, in refs.~\cite{Casini:2013rba, Casini:2014aia, Moitra:2018lxn} it was shown that, in the continuum limit, suitably regularized quantities, such as the mutual information or relative entropies, as well as the universal information encoded in the entanglement entropy, do not get contributions from the classical part of eq.~\eqref{general_expression_entanglement_entropy_gauge_theories} (which depends on high-momentum modes localized close to the boundary), and, therefore, are  independent from the choice of boundary operators.

These subtleties in the definition of the entanglement entropy imply that the replica trick itself presents some ambiguities for gauge theories; in particular, working with the lattice regularization in the Feynman path-integral approach, the problem translates in how to treat the entangling surface where different replicas are joined together.

In this work we address one facet of this problem, focusing on the construction of a replica space for gauge theories through a duality transformation. For gauge theories defined in three spacetime dimensions, the latter can be carried out through a Kramers--Wannier transform~\cite{Kramers:1941kn, Kramers:1941zz}, which maps the gauge theory to a spin system. It is well known that such dualization is not straightforward when the gauge theory is defined on a space with a boundary and there exist operators (including, in particular, those defining Gau{\ss}'s law) defined on the boundary. However, it was shown in different works~\cite{Casini:2014aia, Radicevic:2016tlt, Moitra:2018lxn, Lin:2018bud} that, even though the entanglement entropy is not preserved by the duality transformation, its universal part, encoded in the entropic c-function, is exactly mapped in the continuum limit. Note that the investigation of quantum lattice models by means of duality was recently addressed (in the Hamiltonian formalism) in ref.~\cite{Fazza:2022fss}, albeit with a focus on aspects different from the ones of the present work.

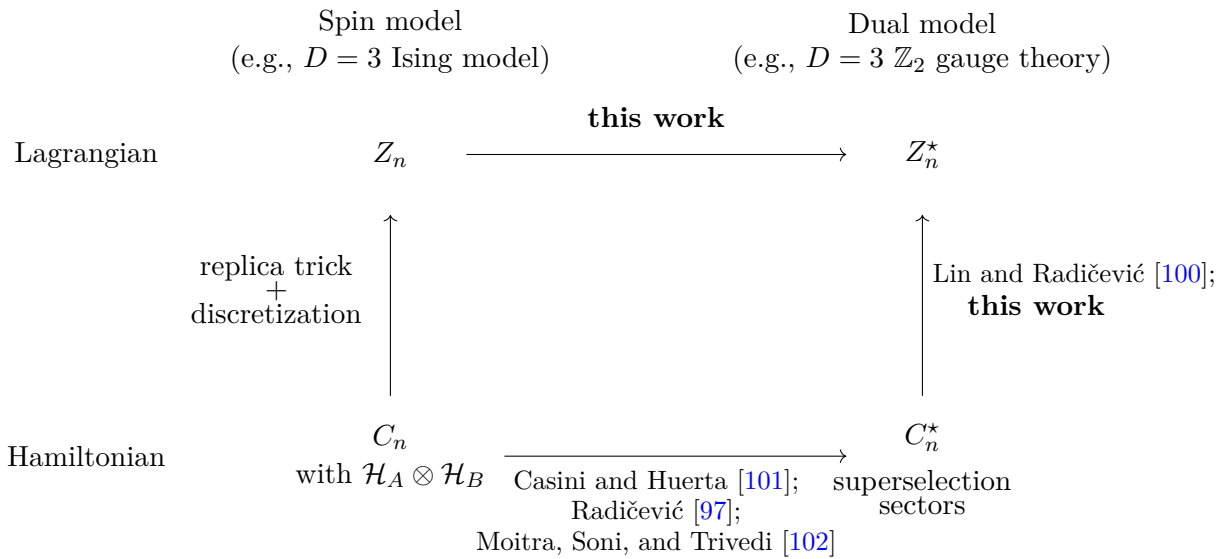
\begin{figure}[t]
\centering
\begin{tikzpicture}[scale=1., every node/.style={scale=1.}]
\draw (0,0) node {Hamiltonian};
\draw (0,4) node {Lagrangian};
\draw (4,5.75) node {Spin model};
\draw (4,5.25) node {(e.g., $D=3$ Ising model)};
\draw (11,5.75) node {Dual model};
\draw (11,5.25) node {(e.g., $D=3$ $\Z_2$ gauge theory)};
\draw (4,.25) node {$C_n$};
\draw (4,-.25) node {with $\mathcal{H}_A\otimes \mathcal{H}_B$};
\draw[->] (4,.8)--(4,3.2);
\draw (4,4) node {$Z_n$};
\draw (2.5, 2.5) node {replica trick};
\draw (2.5, 2.2) node {$+$};
\draw (2.5, 1.9) node {discretization};
\draw[->] (5,4)--(10,4);
\draw (7.5,4.5) node {\textbf{this work}};
\draw (11,4) node {$Z_n^\star$};
\draw[->] (5.5,0) -- (10,0);
\draw (11,0.25) node {$C_n^\star$};
\draw (11,-.35) node {superselection};
\draw (11, -.65) node {sectors};
\draw (7.5, -.35) node {\small{Casini and Huerta~\cite{Casini:2014aia};}};
\draw (7.5,-.75) node {\small{Radi\v{c}evi\'c~\cite{Radicevic:2016tlt};}};
\draw (7.5, -1.15) node {\small{Moitra, Soni, and Trivedi~\cite{Moitra:2018lxn}}};
\draw[->] (11,.8) -- (11,3.2);
\draw (12.5,2.4) node {\hspace{1cm}\small{Lin and Radi\v{c}evi\'c~\cite{Lin:2018bud};}};
\draw (12.5,2.) node {\textbf{this work}};
\end{tikzpicture}
\caption{Starting from a spin model in the Hamiltonian formalism, one can derive the replica geometry of the dual theory following two different paths, which yield equivalent physical results. One consists in discretizing the replica space first, and then performing the duality transformation in the Lagrangian formulation: this is worked out in this article in section~\ref{sec:duality_transformation}. Alternatively, one can transform the spin system at the Hamiltonian level (effects of the duality transformation on universal information encoded in the entanglement entropy were discussed in refs.~\cite{Casini:2014aia, Radicevic:2016tlt, Moitra:2018lxn}), introduce the replicas as described in ref.~\cite{Lin:2018bud} and discretize the resulting system. This approach is discussed in this work in section~\ref{sec:connection_with_the_Hamiltonian_formalism}.}
\label{fig:diagram_of_this_work}
\end{figure}

In the first part of this work, we explicitly perform the duality transformation starting from a spin model defined on a lattice in three spacetime dimensions with an arbitrary positive integer number of replicas, to derive the corresponding geometry for the dual gauge theory. As we discuss in what follows, the physical results we obtain from our construction are purely geometric, namely they do not depend on the degrees of freedom of the theory (nor on the details of how they are assigned to $A$ or $B$); rather, the duality transformation gives us a well-defined prescription on how to implement a conical singularity on the lattice. The logical scheme of how the direct and the dual formulation of a theory are connected with each other, in the Hamiltonian and in the Lagrangian formalisms, is summarized in figure~\ref{fig:diagram_of_this_work} (which also includes the references to previous literature, in which the various aspects discussed herein were studied).

Then, in the second part of this work, we make use of the Kramers--Wannier duality in a complementary way: by means of Monte Carlo simulations, we calculate the entropic c-function of the second R\'enyi entropy of the three-dimensional Ising model, which is dual to the $\Z_2$ gauge theory. This allows one to study the entanglement content of the ground state of a strongly coupled gauge theory and to test the predictions that were derived from the gauge/gravity correspondence in ref.~\cite{Klebanov:2007ws} in a particularly simple (albeit non-trivial) system. The results we obtain are consistent with such predictions and with similar recent studies in low-dimensional gauge theories~\cite{Florio:2023mzk}.

The structure of the article is the following. In section~\ref{sec:duality_transformation} we review the generalities about the duality transformation, and discuss its formulation in the presence of replicas; we mostly focus on the case of the Ising model, considering both the case of two and three spacetime dimensions, but we also discuss the extension of our results to other gauge theories, including $\U(1)$ and $\Z_N$ gauge theories (and briefly comment on the generalization to higher dimensions and to non-Abelian gauge theories). Section~\ref{sec:connection_with_the_Hamiltonian_formalism} illustrates the connection between the Hamiltonian and the Lagrangian formulations in the presence of replicas, with a detailed discussion of the $(1+1)$-dimensional Ising model case study, while section~\ref{sec:monte_carlo} is devoted to a Monte Carlo study of the entropic c-function of the three-dimensional $\Z_2$ gauge theory. Finally, in section~\ref{sec:conclusions} we discuss our findings and comment on possible generalizations thereof. The appendix contains a discussion of the r\^ole of boundary conditions in the duality transformation (section~\ref{app:topology}), a study of the dual replica space for other two-dimensional spin models (section~\ref{app:duality_ZN_U1_2d}), a derivation of the spin theory dual to the three-dimensional $\U(1)$ gauge theory (section~\ref{app:derivation_dual_U1}), and some technical details about our numerical simulations (section~\ref{app:simulation_details}).

\section{Duality transformation}
\label{sec:duality_transformation}

In this section we discuss the duality transformation for a Euclidean lattice including replicas, with a specific focus on the Ising model in two and in three dimensions. We first review the generalities about the Kramers--Wannier duality in subsection~\ref{subsec:review_of_the_duality_transformation}, then in subsection~\ref{subsec:duality_in_the_presence_of_replicas} we present the detailed derivation of the duality transformation for a replica geometry, also commenting on the physical and unphysical contributions to the R\'enyi entropies defined in the direct and in the dual formulation. Finally, we briefly discuss the generalization to other Abelian models in subsection~\ref{subsec:U1_and_ZN_models} and the case of higher dimensions and non-Abelian theories in subsection~\ref{subsec:higher_dimensions_and_non-Abelian_theories}.

\subsection{Review of the duality transformation}
\label{subsec:review_of_the_duality_transformation}

Let us first review the Kramers--Wannier duality transformation, following ref.~\cite{Savit:1979ny}; for simplicity, we discuss the case of the Ising model defined on an isotropic square (in $D=2$ dimensions) or cubic (for $D=3$) lattice with $|\Lambda|$ sites. In what follows, we consider a lattice of arbitrarily large extent in all directions, so that the choice of boundary conditions is immaterial; the case of a finite-extent lattice, with either periodic or antiperiodic boundary conditions in each of the main directions, will be discussed in the appendix~\ref{app:topology}.

Starting from the partition function of the Ising model
\begin{align}
Z(\beta) = \sum_{\{\sigma\}}\exp(\beta\sum_{i}\sum_{\mu}\sigma_i\sigma_{i+\hat{\mu}}),
\label{partition_function_Ising_2d}
\end{align}
where the variables $\sigma_i \in \{1, -1\}$ are defined on the sites of the lattice, and $\mu$ ranges over the main directions of the lattice, by defining the function
\begin{align}
\label{Ck_beta_definition}
C_k(\beta)=\cosh(\beta) \cdot \left\{ 1+k[\tanh (\beta) -1) ]\right\} 
\qquad \mbox{for $k \in \{0,1\}$}
\end{align}
(which equals $\cosh\beta$ for $k=0$, and $\sinh\beta$ for $k=1$), and introducing integer-valued variables $k_{i,\mu}$, associated with the oriented bonds between nearest-neighbor sites on the lattice, eq.~\eqref{partition_function_Ising_2d} can be recast in the form
\begin{align}
Z(\beta) = \sum_{\{\sigma\}}\prod_{i,\mu}\sum_{k_{i,\mu} = 0,1} C_{k_{i,\mu}}(\beta)(\sigma_i\sigma_{i+\hat{\mu}})^{k_{i,\mu}}.
\end{align}
By rearranging the product of $\sigma_i$ terms to isolate each of them, one obtains 
\begin{align}
Z(\beta) = \sum_{\{k\}}\prod_{i,\mu}C_{k_{i,\mu}}(\beta)\sum_{\sigma_i}\sigma_i^{\sum_{i}k_{i,\mu}} = \sum_{\{k\}}\prod_{i,\mu}C_{k_{i,\mu}}(\beta)2\delta_2\left(\sum_{i}k_{i,\mu}\right),
\label{Kronecker_delta_mod_2}
\end{align}
where $\sum_{i}k_{i,\mu}$ denotes the sum over the $k_{i,\mu}$ on all links touching site $i$, and $\delta_2$ is the Kronecker delta modulo~$2$.

The constraint expressed by the latter can be enforced by means of a (discretized) Bianchi identity, by introducing a dual lattice, whose sites are displaced by half a lattice spacing in each direction with respect to the sites of the original lattice. To this purpose, let us first define
\begin{align}
\label{dual_coupling}
\beta^\star = -\frac{1}{2}\ln\tanh\beta.
\end{align}

In $D=2$, assigning $s_i\in\{1, -1\}$ variables to the sites of the dual lattice, the constraint in eq.~\eqref{Kronecker_delta_mod_2} is then satisfied by setting
\begin{align}
k_{i,\mu} = \frac{1 - s_i s_{i+\hat{\nu}}}{2}, \qquad \mbox{with $\nu \neq \mu$}.
\label{solution_of_delta_constraint_2d}
\end{align}
Note, however, that the mapping between configurations of the $s$ variables and configurations of the $k$ variables is surjective, but not injective: configurations of the $s$ variables that differ by an overall sign flip correspond to the same configuration of the $k$ variables. Taking into account this overcounting by a factor $2$, and using eq.~\eqref{solution_of_delta_constraint_2d} in eq.~\eqref{Ck_beta_definition}, the partition function takes the form 
\begin{align}
\label{dual_Ising_2d}
Z(\beta) = \frac{1}{2}[\sinh (2\beta^\star)]^{-\lam}\sum_{\{s\}}\prod_{i,\mu} \exp(\beta^\star s_i s_{i+\hat{\mu}}) = \frac{1}{2}[\sinh (2\beta^\star)]^{-\lam} Z(\beta^\star).
\end{align}
Equation~\eqref{dual_Ising_2d} shows that in two dimensions the partition function of the Ising spin model, evaluated at coupling $\beta$, coincides (up to a trivial normalization factor) with the partition function of the same model, evaluated at the dual coupling $\beta^\star$.

For the $D=3$ Ising model, instead, eq.~\eqref{Kronecker_delta_mod_2} can be rewritten by defining a set of variables $U_{i,\mu}\in\{1,-1\}$ on the links of the dual lattice, and setting
\begin{align}
k_{i,\mu} = \frac{1-U_{i,\nu}U_{i+\hat{\nu},\lambda}U_{i+\hat{\lambda},-\nu}U_{i,-\lambda}}{2}, \qquad \mbox{with $\mu$, $\nu$ and $\lambda$ all different}.
\label{solution_of_delta_constraint_3d}
\end{align}
With manipulations analogous to those for the $D=2$ case, and denoting the product of four $U_{i,\mu}$ variables around a plaquette as $U_{\Box}$, one obtains
\begin{align}
Z(\beta) = \frac{\sum_{\{U\}}\prod_{\Box} \exp(\beta^\star U_\Box)}{2^{-\frac{\lam}{2}-\Ng}[\sinh (2\beta^\star)]^{-\frac{3\lam}{2}}} = \frac{Z^{\mbox{\tiny{gauge}}}(\beta^\star)}{2^{\frac{\lam}{2}+\Ng}[\sinh (2\beta^\star)]^{\frac{3\lam}{2}}}.
\label{dual_theory_to_Ising_3d}
\end{align}
As in $D=2$, the mapping between configurations of the $U$ variables and configurations of the $k$ variables is not a bijective one. In this case, each configuration of the $k$'s corresponds to $2^{\Ng}$ configurations of the $U$'s, that differ from each other only by gauge transformations (i.e., by flips of the signs of the $U$'s on all of the links touching a site): here $\Ng$ denotes the number of different $U$ variables that can be set to a chosen value (say, $1$) purely by gauge transformations, i.e., the number of links in a maximal tree on the lattice; note that $\Ng$ is strictly smaller than $\lam$. Equation~\eqref{dual_theory_to_Ising_3d} shows that, up to an overall normalization factor, the partition function of the $D=3$ Ising model evaluated at coupling $\beta$ equals the partition function of the three-dimensional $\Z_2$ gauge theory, evaluated at coupling $\beta^\star$.

\subsection{Duality in the presence of replicas}
\label{subsec:duality_in_the_presence_of_replicas}

\begin{figure}[t]
\centering
\begin{tikzpicture}[scale=1.2, every node/.style={scale=1.2}]
\draw[step = 1cm, gray, thin] (-.9, -.9) grid (3.9, 3.9);
\foreach \x in {0,...,3}{
  \foreach \y in {0,...,3}{
    \fill[color=black] (\x, \y) circle (.1);
  }
}
\draw[step = 1cm, gray, thin] (5.1, -.9) grid (9.9, 3.9);
\foreach \x in {6,...,9}{
  \foreach \y in {0,...,3}{
    \fill[color=black] (\x, \y) circle (.1);
  }
}
\draw[snake= coil, segment aspect = 0, color=red] (.5,1.5) -- (2.5,1.5);
\draw[snake= coil, segment aspect = 0, color=red] (6.5,1.5) -- (8.5,1.5);
\draw[ultra thick, color=green!70!black] (1,1.9) -- (1,1.5);
\draw[ultra thick, color=green!70!black] (2,1.9) -- (2,1.5);
\draw[ultra thick, color=green!70!black] (7,1.5) -- (7,1.1);
\draw[ultra thick, color=green!70!black] (8,1.5) -- (8,1.1);
\draw[->] (-2,-1) -- (-2,0);
\draw[->] (-2,-1) -- (-1,-1);
\draw (-1.7,0) node {$\hat{0}$};
\draw (-1, -0.7) node {$\hat{1}$};
\draw[color=gray, dashed] (.5,-1) -- (.5,4);
\draw[color=gray, dashed] (2.5,-1) -- (2.5,4);
\draw[color=gray, dashed] (6.5,-1) -- (6.5,4);
\draw[color=gray, dashed] (8.5,-1) -- (8.5,4);
\draw (1.5,4) node {$A$};
\draw (-.3,4) node {$B$};
\draw (3.3,4) node {$B$};
\draw (7.5,4) node {$A$};
\draw (5.7,4) node {$B$};
\draw (9.3,4) node {$B$};
\end{tikzpicture}
\caption{Two-dimensional replica space. The vertical direction is the Euclidean-time direction, while the horizontal one is the spatial one. The red wavy line represents the cut; green links connect spins in different replicas, therefore the periodicity of the lattice in the subsystem $A$ is $n$ times the periodicity in the subsystem $B$.}
\label{fig:replica_space_cut}
\end{figure}
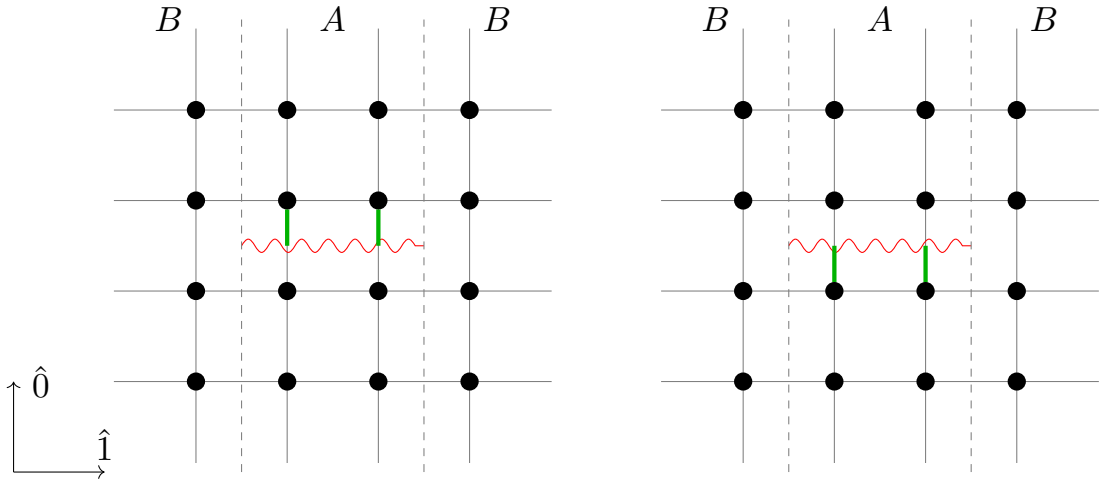

Let us now discuss the generalization of the duality in the presence of a set of replicas. Note that our presentation in this subsection is strictly a geometric one, while section~\ref{sec:connection_with_the_Hamiltonian_formalism} contains a more detailed discussion about how a more algebraic approach can be mapped to the construction that we present here, and how the various degrees of freedom of the theory are encoded in this setup.

\begin{figure}[t]
\centering
\begin{tikzpicture}[scale=1.2, every node/.style={scale=1.2}]
\draw (0,-2) -- (0,2);
\draw (-2,0) -- (2,0);
\fill[color=black] (0, 0) circle (.1);
\draw[gray, thin, dashed] (-2,-1) -- (2,-1);
\draw[gray, thin, dashed] (-2,1) -- (2,1);
\draw[gray, thin, dashed] (-1,-2) -- (-1,2);
\draw[gray, thin, dashed] (1,-2) -- (1,2);
\fill[color=gray] (1, 1) circle (.1);
\fill[color=red] (-1, 1) circle (.1);
\fill[color=gray] (1, -1) circle (.1);
\fill[color=gray] (-1, -1) circle (.1);
\draw (-1.4,1.4) node {$s_{j+\hat{0}}^{(1)}$};
\draw (1.5,1.4) node {$s_{j+\hat{1}+\hat{0}}^{(1)}$};
\draw (1.4,-1.3) node {$s_{j+\hat{1}}^{(1)}$};
\draw (-1.3,-1.3) node {$s_j^{(1)}$};

\draw (2.3,0) node {$k_{i,1}^{(1)}$};
\draw (0,-2.3) node {$k_{i-\hat{0},0}^{(1)}$};
\draw (-2.3,0) node {$k_{i-\hat{1},1}^{(1)}$};
\draw (0,2.3) node {$k_{i,0}^{(n)}$};
\draw (.2,-.3) node {$i$};
\draw (-.3,.3) node {$k_{i,0}^{(1)}$};
\draw[snake= coil, segment aspect = 0, color = red] (-1,1) -- (-.7,.7) -- (2,0.7);
\draw (6,-2) -- (6,2);
\draw (4,0) -- (8,0);
\fill[color=black] (6, 0) circle (.1);
\draw[gray, thin, dashed] (4,-1) -- (8,-1);
\draw[gray, thin, dashed] (4,1) -- (8,1);
\draw[gray, thin, dashed] (5,-2) -- (5,2);
\draw[gray, thin, dashed] (7,-2) -- (7,2);
\fill[color=gray] (7, 1) circle (.1);
\fill[color=red] (5, 1) circle (.1);
\fill[color=gray] (7, -1) circle (.1);
\fill[color=gray] (5, -1) circle (.1);
\draw (-1.4+6,1.4) node {$s_{j+\hat{0}}^{(2)}$};
\draw (1.5+6,1.4) node {$s_{j+\hat{0}+\hat{1}}^{(2)}$};
\draw (1.4+6,-1.3) node {$s_{j+\hat{1}}^{(2)}$};
\draw (-1.3+6,-1.3) node {$s_j^{(2)}$};
\draw (2.3+6,0) node {$k_{i,1}^{(2)}$};
\draw (6,-2.3) node {$k_{i-\hat{0},0}^{(2)}$};
\draw (-2.3+6,0) node {$k_{i-\hat{1},1}^{(2)}$};
\draw (6,2.3) node {$k_{i,0}^{(1)}$};
\draw (.2+6,-.3) node {$i$};
\draw (-.3+6,.3) node {$k_{i,0}^{(2)}$};
\draw[snake= coil, segment aspect = 0, color = red] (5,1) -- (-.7+6,.7) -- (2+6,0.7);
\end{tikzpicture}
\caption{Structure of the dual lattice close to a branch point (red dot).}
\label{fig:duality_close_to_the_cut_Ising_2d}
\end{figure}
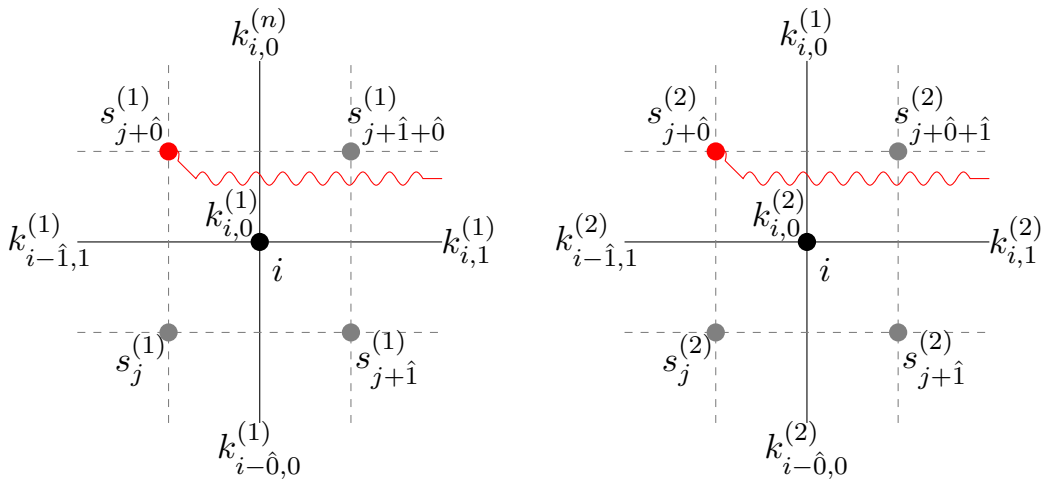

The geometry we are interested in is the one of $n$ replicas connected through a cut, and let us first consider the two-dimensional case, as sketched in figure~\ref{fig:replica_space_cut}. In a replica geometry, the expression~\eqref{Kronecker_delta_mod_2} still holds, since its derivation is only based on the nearest-neighbor nature of the interaction and is independent from the geometry of the lattice. The difference in the duality transformation is in the representation of the $k_{i,\mu}$ variables that satisfy the constraint; clearly, the introduction of a cut does not affect the duality transformation for links not crossed by the cut, since the transformation only involves nearest-neighbor spins in the dual lattice and the links in the direct lattice separating them. Moreover, the duality transformation remains unchanged also for links crossed by the cut that are far from the cut boundary.\footnote{This is related to the fact that the exact location of the cut is irrelevant.} The only difference arises at the edges of the cut; in particular, in two dimensions the dual lattice has sites exactly at (the lattice analogue of) the branch points of the replica space. Consider, for example, the cartoon in figure~\ref{fig:duality_close_to_the_cut_Ising_2d}, where two of the $n$ replicas are depicted: the equation expressing the constraint associated with the site $i$ of the direct lattice, in the replica labeled $1$, is
\begin{align}
k_{i,0}^{(1)} + k_{i,1}^{(1)} + k_{i-\hat{0},0}^{(1)} + k_{i-\hat{1},1}^{(1)} = 0.
\end{align}
Proceeding as in subsection~\ref{subsec:review_of_the_duality_transformation}, and denoting the site of the dual lattice in the lower left corner of the plaquette surrounding $i$ as $j$, one gets
\begin{align*}
k_{i,0}^{(1)} = \frac{1 - s^{(2)}_{j+\hat{0}}s^{(2)}_{j+\hat{0}+\hat{1}}}{2}, \quad
k_{i,1}^{(1)} = \frac{1 - s^{(2)}_{j+\hat{0}+\hat{1}}s^{(1)}_{j+\hat{1}}}{2}, \\
k_{i-\hat{0},0}^{(1)} = \frac{1 - s^{(1)}_{j}s^{(1)}_{j+\hat{1}}}{2},\quad
k_{i-\hat{1},1}^{(1)} = \frac{1 - s^{(1)}_{j}s^{(1)}_{j+\hat{0}}}{2},
\end{align*}
and similar relations on the other replicas. Clearly, the constraint is satisfied if and only if the spins defined on the branch-point sites of all replicas coincide: $s^{(1)}_{j+\hat{0}} = s^{(2)}_{j+\hat{0}} = \dots = s^{(n)}_{j+\hat{0}}$, meaning that the sites lying on (the lattice counterpart of) the conical singularity of the replica manifold belong to all replicas, hence on a square lattice they have $4n$ nearest neighbors. Thus, the partition function of the system with $n$ replicas can be written as
\begin{align}
Z_n(\beta) = 2^{n\lam -1}\sum_{\{s\}}\sum_{r=0}^n \prod_{i,\mu}C_{\frac{1-s^{(r)}_i s^{(r)}_{i+\hat{\mu}}}{2}}(\beta) = \frac{Z_n^\star(\beta^\star)}{2[\sinh (2\beta^\star)]^{n\lam}},
\label{duality_eq_replica_2d}
\end{align}
where $Z_n^\star(\beta^\star)$ denotes the partition function of a system in which the spins defined on the branch points have $4n$ nearest neighbors. In fact, the enlarged number of nearest neighbors close to a branch point has been pointed out before~\cite{Cardy:2010zs}. Note that for a finite lattice with specific boundary conditions eq.~\eqref{duality_eq_replica_2d} is expected to get contributions from different topological sectors, as discussed in detail in section~\ref{app:topology} of the appendix.

Using eq.~\eqref{dual_Ising_2d} and eq.~\eqref{duality_eq_replica_2d}, one can derive an explicit relation between the R\'enyi entropies in the direct and in the dual formulation of the model:
\begin{align}
\frac{Z_n(\beta)}{Z^n(\beta)} = 2^{n-1}\frac{Z_n^\star(\beta^\star)}{(Z^\star(\beta^\star))^n},
\end{align}
which leads to
\begin{align}
S_n^\star(\beta^\star) = \ln 2 + S_n(\beta),
\label{Renyi_entropy_under_duality_2d}
\end{align}
therefore the R\'enyi entropies evaluated in the direct and in the dual formulation are not equal: note, however, that their difference can be traced back to the factor $\frac{1}{2}$ arising from the change of variables from $k$ to $s$ and therefore it is related to the lattice degrees of freedom we are using to describe the model, which is clearly an unphysical piece of information. By contrast, the entropic c-function, which is a quantity encoding physical information, is unaffected by the additive constant $\ln 2$ and we get
\begin{align}
C^\star_n(l,\beta^\star) = C_n(l,\beta), 
\end{align}
where $l$ denotes the length of the cut.

\begin{figure}[t]
\centering
\begin{tikzpicture}[scale=1.4, every node/.style={scale=1.4}]
\fill[color=red!50!white, fill opacity=.5] (-1.5,.7) to[out=-30, in=180] (1,.4) -- (1.5,.9) to[out=180, in=-30] (-1,1.2) -- (-1.5,.7);
\draw (-1.5,-1.5)--(1.5,1.5);
\draw (-2,0)--(2,0);
\draw (0,-2)--(0,2);
\draw[dashed] (.5,-1.3)--(.5,.7); 
\draw[dashed] (-1.5,-1.3)--(.5,-1.3);
\draw[dashed] (-1.5,-1.3)--(-1.5,.7); 
\draw[dashed] (-1.5,.7)--(.5,.7);
\draw[dashed] (.5,-1.3)--(1,-.8);
\draw[dashed] (-1.5,-1.3)--(-1,-.8);
\draw[dashed] (.5,.7)--(1,1.2);
\draw[red,very thick] (-1.5,.7)--(-1,1.2);
\draw[dashed] (-1,-.8)--(1,-.8);
\draw[dashed] (-1,1.2)--(1,1.2);
\draw[dashed] (-1,-.8)--(-1,1.2);
\draw[dashed] (1,-.8)--(1,1.2);
\fill[color=black] (0,0) circle (.1);
\fill[color=black] (-1.5,-1.3) circle (.1);
\fill[color=red] (0,.7) circle (.07);
\draw (.3,-.3) node {$\sigma^{(1)}_i$};
\draw (-1.8,-1.3) node {$j$};
\draw (-1,-1.7) node {$k^{(1)}_{i-\hat{2},2}$};
\draw (1.8,1.6) node {$k^{(1)}_{i,2}$};
\draw (1.5,-.3) node {$k^{(1)}_{i,1}$};
\draw (-2,-.3) node {$k^{(1)}_{i-\hat{1},1}$};
\draw (.5,-1.7) node {$k^{(1)}_{i-\hat{0},0}$};
\draw (-.4,.25) node {$k^{(1)}_{i,0}$};
\draw (-.4,1.6) node {$k^{(n)}_{i,0}$};
\draw[->] (-2.2-.4,-1.5)--(-1.7-.4,-1.5);
\draw[->] (-2.2-.4,-1.5)--(-1.95-.4,-1.25);
\draw[->] (-2.2-.4,-1.5)--(-2.2-.4,-1);
\draw (-2.4-.4,-1.2) node[scale = .8] {$\hat{0}$};
\draw (-1.8-.4,-1.2) node[scale = .8] {$\hat{2}$};
\draw (-1.75-.4,-1.7) node[scale = .8] {$\hat{1}$};
%
%
\fill[color=red!50!white, fill opacity=.5] (-1.5+5,.7) to[out=-30, in=180] (1+5,.4) -- (1.5+5,.9) to[out=180, in=-30] (-1+5,1.2) -- (-1.5+5,.7);
\draw (-1.5+5,-1.5)--(1.5+5,1.5);
\draw (-2+5,0)--(2+5,0);
\draw (0+5,-2)--(0+5,2);
\draw[dashed] (.5+5,-1.3)--(.5+5,.7); 
\draw[dashed] (-1.5+5,-1.3)--(.5+5,-1.3);
\draw[dashed] (-1.5+5,-1.3)--(-1.5+5,.7); 
\draw[dashed] (-1.5+5,.7)--(.5+5,.7);
\draw[dashed] (.5+5,-1.3)--(1+5,-.8);
\draw[dashed] (-1.5+5,-1.3)--(-1+5,-.8);
\draw[dashed] (.5+5,.7)--(1+5,1.2);
\draw[red,very thick] (-1.5+5,.7)--(-1+5,1.2);
\draw[dashed] (-1+5,-.8)--(1+5,-.8);
\draw[dashed] (-1+5,1.2)--(1+5,1.2);
\draw[dashed] (-1+5,-.8)--(-1+5,1.2);
\draw[dashed] (1+5,-.8)--(1+5,1.2);
\fill[color=black] (0+5,0) circle (.1);
\fill[color=black] (-1.5+5,-1.3) circle (.1);
\fill[color=red] (0+5,.7) circle (.07);
\draw (.3+5,-.3) node {$\sigma^{(2)}_i$};
\draw (-1.8+5,-1.3) node {$j$};
\draw (-1+5,-1.7) node {$k^{(2)}_{i-\hat{2},2}$};
\draw (1.8+5,1.6) node {$k^{(2)}_{i,2}$};
\draw (1.5+5,-.3) node {$k^{(2)}_{i,1}$};
\draw (-2+5,-.3) node {$k^{(2)}_{i-\hat{1},1}$};
\draw (.5+5,-1.7) node {$k^{(2)}_{i-\hat{0},0}$};
\draw (-.4+5,.25) node {$k^{(2)}_{i,0}$};
\draw (-.4+5,1.6) node {$k^{(1)}_{i,0}$};
\end{tikzpicture}
\caption{Duality transformation in a three-dimensional replica space. The solid line is the entangling surface along the links of the dual lattice, the red region is the cut, the red dot the intersection between the link between the sites $i$ and $i+\hat{0}$ and the cut.}
\label{fig:duality_Ising_3d}
\end{figure}
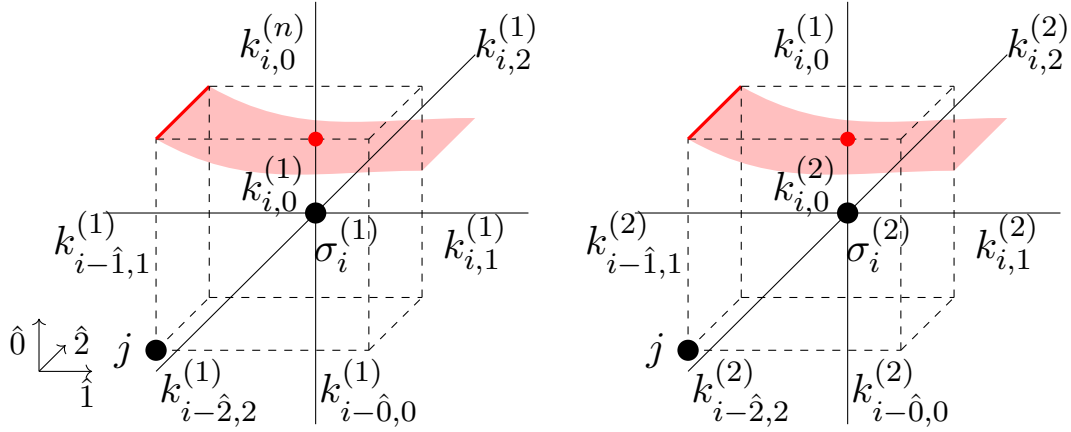

It is straightforward to carry out the dualization in the presence of replicas for the three-dimensional Ising model, too. In this case (see the sketch in figure~\ref{fig:duality_Ising_3d}) the ``cut'' is actually a surface, and the conical singularity is a line where the cut ends. As for the two-dimensional model, the way the duality maps degrees of freedom in the direct and in the dual formulation is not affected by the presence of conical singularities, for spins far enough from them. Consider, for example, the site $i$ in the direct lattice shown in figure~\ref{fig:duality_Ising_3d}; in the first replica, the Kronecker-delta constraint is of the form
\begin{align}
k^{(1)}_{i,0} + k^{(1)}_{i-\hat{0},0} + k^{(1)}_{i,1} + k^{(1)}_{i-\hat{1},1} + k^{(1)}_{i,2} + k^{(1)}_{i-\hat{2},2} = 0.
\label{local_constraint_equation_replica_3d}
\end{align}
Repeating the construction carried out in subsection~\ref{subsec:review_of_the_duality_transformation}, one obtains
\begin{align*}
k^{(1)}_{i,0} = \frac{1 - U^{(2)}_{j+\hat{0},1} U^{(2)}_{j+\hat{0}+\hat{1},2} U^{(2)}_{j+\hat{0}+\hat{2},1} U^{(2)}_{j+\hat{0},2}}{2}, \quad k^{(1)}_{i-\hat{0},0} = \frac{1 - U^{(1)}_{j,1} U^{(1)}_{j+\hat{1},2} U^{(1)}_{j+\hat{2},1} U^{(1)}_{j,2}}{2},\\
k^{(1)}_{i,1} = \frac{1 - U^{(1)}_{j+\hat{1},2} U^{(1)}_{j+\hat{1}+\hat{2},0} U^{(2)}_{j+\hat{1}+\hat{0},2} U^{(1)}_{j+\hat{1},0}}{2}, \quad k^{(1)}_{i-\hat{1},1} = \frac{1 - U^{(1)}_{j,0} U^{(1)}_{j+\hat{0},2} U^{(1)}_{j+\hat{2},0} U^{(1)}_{j,2}}{2},\\
k^{(1)}_{i,2} = \frac{1 - U^{(1)}_{j+\hat{2},1} U^{(1)}_{j+\hat{2}+\hat{1},0} U^{(2)}_{j+\hat{2}+\hat{0},1} U^{(1)}_{j+\hat{2},0}}{2}, \quad k^{(1)}_{i-\hat{2},2} = \frac{1 - U^{(1)}_{j,1} U^{(1)}_{j+\hat{1},0} U^{(2)}_{j+\hat{0},1} U^{(1)}_{j,0}}{2}.
\end{align*}
Inspecting the previous equations we can notice that gauge fields $U^{(1)}_{j+\hat{0},2}$ and $U^{(2)}_{j+\hat{0},2}$, the ones lying on the conical singularity, appear only once in eq.~\eqref{local_constraint_equation_replica_3d}. Therefore the constraint is satisfied if $U^{(1)}_{j+\hat{0},2} = U^{(2)}_{j+\hat{0},2}$. This is true for all gauge fields on the entangling surface. Therefore in the dual replica space the gauge fields lying on the conical singularity are common to all replicas and they appear in the action in $4n$ different plaquette terms, $4$ for each replica. Note that this construction can be performed independently for each link of the entangling surface, and is independent from the shape of the latter. 

The presence of gauge fields shared among different replicas leads to some constraints on the gauge transformations that leave the physical properties of the system invariant. In particular, there exist plaquette terms that can be schematically written as $U_{\mbox{\tiny{edge}}}U^{(r)}_{\mbox{\tiny{staple}}}$, where $U_{\mbox{\tiny{edge}}}$ is the gauge field on the conical singularity and $U^{(r)}_{\mbox{\tiny{staple}}}$ is a staple in replica $r$ forming a plaquette together with $U_{\mbox{\tiny{edge}}}$. The presence of such terms require that gauge transformations performed on the sites along the entangling surface to act on all replicas. More precisely, a gauge transform on a site on the entangling surface acts upon $(4n + 2)$ gauge fields: the $4n$ links orthogonal to the conical singularity attached to that point, and the two links along the entangling surface shared by all replicas. This restriction on the possible gauge transformations in a replica space was also noted in the context of $\SU(N)$ gauge theories~\cite{Soni:2016ogt, Jokela:2022fvh}.

An interesting implication of the geometry of the entangling surface is that it is possible to define Wilson loops belonging to different replicas, passing through sites on the conical singularity. These loops are closed in the full replica space, but looking at the single replica they appear to be open, with both endpoints on the entangling surface.

We can now write down the expression of the duality transformation in replica space: for a connected entangling surface made of $|\partial A|$ lattice sites, we have
\begin{align}
Z_n(\beta) = 2^{(n-1)(|\partial A|-1)}2^{-\frac{\lam n}{2}-\Ng n}(\sinh 2\beta^\star)^{-\frac{3\lam n}{2}} Z_n^{\Z_2}(\beta^\star),
\label{duality_eq_enhanced_nn_3d}
\end{align}
where $Z_n^{\Z_2}$ is the partition function of the gauge model in the geometry that we discussed. Note that the factor $2^{(n-1)(|\partial A| - 1)}$ is a correction related to the number of gauge transforms which can be performed in this geometry. As for the two-dimensional case, also here for a sufficiently large system one can neglect the r\^ole of boundary conditions and the non-trivial topological implications they entail.

Combining eq.~\eqref{dual_theory_to_Ising_3d} with eq.~\eqref{duality_eq_enhanced_nn_3d}, one can derive the expression for the R\'enyi entropy of the dual gauge theory
\begin{align}
S^{\Z_2}_n(\beta^\star) = (|\partial A|-1)\ln 2 + S_n(\beta).
\label{Renyi_entropy_under_duality_3d}
\end{align}
The difference between the R\'enyi entropies in the gauge and in the spin theories is the presence of an ultraviolet-divergent term proportional to the length of the entangling surface, therefore contributing to the area law. This term makes the R\'enyi entropy of the gauge theory larger than the one of the Ising model, as was already pointed out in the literature~\cite{Tagliacozzo:2010vk}. The origin of this difference stems from the constraint on the number of possible gauge transforms that can be defined in a system with replicas, and results into a multiplicative constant affecting the partition function for the dual gauge theory, which obviously does not carry any physical information. Indeed, one can also note that the term $|\partial A|-1$ depends on the gauge choice at the entangling surface, as was also discussed in refs.~\cite{Casini:2013rba, Aoki:2015bsa, Aoki:2016lma}. Like in the $D=2$ case, physical information can be obtained from a suitably defined entropic c-function, in which the area-law term cancels out; in $D=3$ if the length of the entangling surface does not depend on the size of the region $A$, as in the slab geometry, the entropic c-function can be simply defined according to eq.~\eqref{definition_entropic_c-function_slab}, and one gets
\begin{align}
C^{\Z_2}_n(l,\beta^\star) = C_n(l,\beta).
\label{entropic_c-function_under_duality_3d}
\end{align} 

Even though the first term appearing in eq.~\eqref{Renyi_entropy_under_duality_3d} is unphysical, it is worth inspecting it further. In particular, in the high-temperature limit of the spin model, $\beta\rightarrow 0$ (which corresponds to the weak-coupling limit $\beta^\star\rightarrow\infty$ of the gauge theory), the R\'enyi entropy reads
\begin{align}
S_n(\beta = 0) = 0, \qquad S^{\Z_2}_n(\beta^\star\rightarrow\infty) = (|\partial A|-1)\ln 2,
\label{topological_EE_from_duality}
\end{align}
which is the result obtained with different definitions of entanglement entropy in Hamiltonian lattice gauge theories~\cite{Buividovich:2008gq, Donnelly:2011hn, Casini:2013rba, Radicevic:2014kqa, Aoki:2015bsa, Radicevic:2015sza, Soni:2015yga, Radicevic:2016tlt, Lin:2018bud}; in particular the $-\ln2$ term appearing on the right-hand side of eq.~\eqref{topological_EE_from_duality} corresponds to the topological entanglement entropy~\cite{Kitaev:2005dm,Levin:2006zz}.

It is also worth noting that, in the opposite limit ($\beta\rightarrow \infty$, or $\beta^\star\rightarrow 0$) the result one would na\"{\i}vely obtain is not the same as in the Hamiltonian formalism:
\begin{align}
S_n(\beta\rightarrow\infty) = \ln 2, \qquad S^{\Z_2}_n(\beta^\star = 0) = |\partial A|\ln 2
\end{align}
(which is different from the expected value $S^{\Z_2}_n(\beta^\star = 0) = 0$): this mismatch is simply due to the fact that in the ordered, low-temperature phase of the Ising model it is no longer justified to assume that the topologically inequivalent sectors induced by the different possible boundary conditions can be neglected in the duality transformation. We present a more detailed discussion about this point in section~\ref{app:topology} of the appendix.

Before concluding this subsection, we would like to point out a possible generalization of the duality transformation for the three-dimensional Ising model in the presence of replicas.

So far, we assumed the entangling surface to lie on the lattice dual to the Ising-model lattice, that is, to be placed along the links of the dual gauge theory. However, it is also interesting to analyze a geometry in which the entangling surface lies on the links of the spin model; in this case there are spin degrees of freedom on the conical singularity. From the discussion of the $D=2$ case, we know that such degrees of freedom are shared among all replicas, and have a larger number of nearest neighbors, which here is $(4n + 2)$, of which $4n$ in the directions orthogonal to the entangling surface and $2$ along the conical singularity.

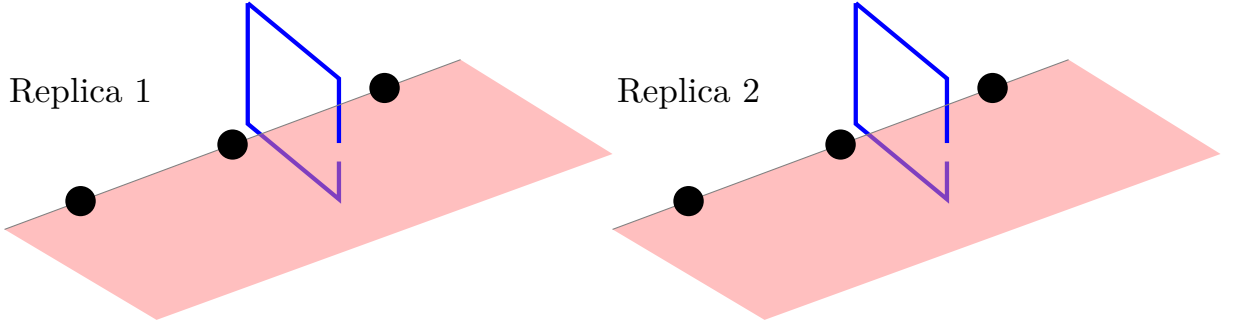
\begin{figure}[t]
\centering
\begin{tikzpicture}[scale=2, every node/.style={scale=1.5}]
\draw[color=blue, ultra thick] (0.6,1.) -- (0.6, .2) -- (1.2, -.3) -- (1.2, -.05);
\fill[color=red!50!white, fill opacity=.5] (-1,-.5) -- (2,.625) -- (3,0) -- (0,-1.1) -- (-1,-.5);
\draw[color=blue, ultra thick] (1.2,.0725) -- (1.2,.5) -- (0.6,1.);
\draw[gray, thin] (-1,-.5) -- (2,.625);
\fill[color=black] (-.5,-.3125) circle (.1);
\fill[color=black] (.5,.0625) circle (.1);
\fill[color=black] (1.5,.4375) circle (.1);
\draw[color=blue, ultra thick] (4.6,1.) -- (4.6, .2) -- (5.2, -.3) -- (5.2, -.05);
\fill[color=red!50!white, fill opacity=.5] (3,-.5) -- (6,.625) -- (7,0) -- (4,-1.1) -- (3,-.5);
\draw[color=blue, ultra thick] (5.2,.0725) -- (5.2,.5) -- (4.6,1.);
\draw[gray, thin] (3,-.5) -- (6,.625);
\fill[color=black] (3.5,-.3125) circle (.1);
\fill[color=black] (4.5,.0625) circle (.1);
\fill[color=black] (5.5,.4375) circle (.1);
\draw (-.5,.4) node[scale=.8] {Replica $1$};
\draw (3.5,.4) node[scale=.8] {Replica $2$};
\end{tikzpicture}
\caption{Central-plaquette geometry. The blue loop crosses the cut (pink surface) and therefore has length $4n$, where $n$ is the number of replicas.}
\label{fig:central_plaquette_geometry}
\end{figure}

As before, let us focus on the duality transformation close to the entangling surface. Consider a link in the direct lattice connecting two spins on the conical singularity, which is shared by all replicas: the dual plaquette orthogonal to that link has an edge, oriented in the Euclidean-time direction, that crosses the cut, therefore the plaquette, before closing on itself, has to loop around the conical singularity $n$ times. The link variable in the direct lattice is therefore represented as
\begin{align}
k_{\mbox{\tiny{edge}}} = \frac{1 - U_{\Box\times n}}{2} ,
\end{align}
where $U_{\Box\times n}$ denotes the closed loop of length $4n$ that winds around the conical singularity. Note that the presence of two such loops of length $4n$ is necessary to satisfy the Kronecker-delta constraint around a point of the direct lattice with an enhanced number of nearest neighbors: for the $4n$ links touching a site on the conical singularity one can make the usual duality transformation according to eq.~\eqref{solution_of_delta_constraint_3d}, but then there are only two link variables to be adjusted, such that in the sum $\sum_i k_{i,\mu}$ all gauge variables appear twice. As a consequence, the two $k$ variables along the conical singularity must be built from a larger number of gauge variables. Note that this geometry was already discussed in ref.~\cite{Chen:2015kfa}, whose authors carried out a strong-coupling calculation of the R\'enyi entropy in this ``central-plaquette'' geometry, where the central plaquettes are the ones encircling the conical singularity, as sketched in figure~\ref{fig:central_plaquette_geometry}.

In this case, the relation between the spin and the gauge partition functions reads
\begin{align}
Z_n(\beta) = 2^{\frac{1}{2}(n-1)|\partial A| -\frac{1}{2}\lam n - \Ng n}(\sinh 2\beta^\star)^{-\frac{3}{2}\lam n + \frac{n-1}{2}|\partial A|}Z_n^{\Z_2}(\beta^\star),
\end{align}
where $Z_n$ and $Z_n^{\Z_2}$ are the partition functions for the spin system and for the gauge model with the geometries discussed in the present section, that are different from the ones appearing in eq.~\eqref{duality_eq_enhanced_nn_3d}. In this central-plaquette geometry, the relation between the R\'enyi entropies associated with the spin model and with the gauge theory reads
\begin{align}
S^{\Z_2}_n(\beta^\star) = - \frac{1}{2}|\partial A|\ln\frac{\sinh(2\beta)}{2} + S_n(\beta),
\end{align}
which is different with respect to eq.~\eqref{Renyi_entropy_under_duality_3d}, 
but, again, the difference consists of unphysical terms and eq.~\eqref{entropic_c-function_under_duality_3d} still holds.

\subsection{$\U(1)$ and $\Z_N$ models}
\label{subsec:U1_and_ZN_models}

The construction described above can be readily generalized to other Abelian gauge theories defined in three spacetime dimensions. In particular, here we will discuss the case of the $\U(1)$ gauge theory (and briefly comment on the generalization to $\Z_N$ gauge theories, too).

The $\U(1)$ gauge theory is defined by the action
\begin{align}
Z^{\U(1)} = \int DU \exp(\beta\sum_\Box \Re U_\Box) ,
\end{align}
where $U_\Box \equiv U_{i,\mu}U_{i+\hat{\mu},\nu}U^\dagger_{i+\hat{\nu},\mu} U^\dagger_{i,\nu}$, and each link variable $U_{i,\mu}$ is an element of the $\U(1)$ group. We can equivalently write $U_{i,\mu} = \exp(i\theta_{i,\mu})$ and directly work with the algebra-valued variables $\theta_{i,\mu}$,
\begin{align}
Z^{\U(1)} = \int D\theta \exp(\beta\sum_{\Box}\cos\theta_\Box),
\end{align}
having defined the discretized exterior derivative of the $\theta$ variables as $\theta_{\Box} \equiv \theta_{i,\mu} + \theta_{i+\hat{\mu},\nu} - \theta_{i+\hat{\nu},\mu} - \theta_{i,\nu}$. The theory admits a dual description in terms of a non-compact, integer-valued spin model, with degrees of freedom $\{\phi\}$; an explicit derivation of the duality is reported in section~\ref{app:derivation_dual_U1} of the appendix. The partition function of the spin model reads
\begin{align}
Z = \sum_{\{\phi\}}\prod_{i,\mu}I_{\Delta_\mu\phi_i}(\beta),
\label{dual_to_U1}
\end{align}
where $I_\nu(\beta)$ is a modified Bessel function of the first kind. Our purpose is to use the duality transformation to derive the (geometric) structure of the replica space of a gauge theory, starting from a spin system. It is therefore useful to reverse the derivation presented in section~\ref{app:derivation_dual_U1} of the appendix to obtain the $\U(1)$ model as the theory dual to the spin system with the partition function in eq.~\eqref{dual_to_U1}, before extending it to a replica geometry.

Using a well-known property of modified Bessel functions of the first kind
\begin{align}
I_n(\beta) = \frac{1}{\pi}\int_0^\pi \dd x e^{\beta\cos x} \cos(nx) = \frac{1}{2\pi}\int_{-\pi}^\pi \dd x e^{\beta\cos x + inx},
\end{align}
the partition function in eq.~\eqref{dual_to_U1} can be rewritten as
\begin{align}
Z & \propto \sum_{\{\phi\}}\prod_{i,\mu}\int D k_{i,\mu} \exp(\beta\cos k_{i,\mu} + i\Delta_\mu\phi_i k_{i,\mu}) \\
&= \int D k \prod_{i,\mu} \exp(\beta \cos k_{i,\mu}) \sum_{\phi_i}\exp(-i\phi_i \sum_i \Delta_\mu k_{i,\mu}) \propto \\
&\propto \int D k \prod_{i,\mu} \exp(\beta\cos k_{i,\mu}) \delta\left(\sum_i \Delta_\mu k_{i,\mu}\right).
\label{derivation_dual_to_dual_to_U1} 
\end{align}
Now $k_{i,\mu}$ is a continuous and periodic link variable and the argument of the Kronecker delta involves the sum over the six links touching site $i$. The constraint is satisfied if one takes
\begin{align}
k_{i,\mu} = \theta_\Box,
\label{solution_delta_U1_3d}
\end{align}
where $\Box$ is the orthogonal plaquette in the dual lattice.

In the presence of replicas, the duality transformation can be generalized along the lines that we discussed for the three-dimensional Ising model; the geometry we consider is the same as in figure~\ref{fig:duality_Ising_3d}, but with $\phi$ variables instead of $\sigma$ ones. The constraint associated to the site $i$ of the left replica is again expressed by eq.~\eqref{local_constraint_equation_replica_3d}. Note that, with the cut placed as in figure~\ref{fig:duality_Ising_3d}, all links along the $0$ direction (both in the direct and in the dual lattice) intersect the cut, and therefore are shared between different replicas. As a consequence, a plaquette in the dual lattice that lies along the $0$ direction has some links in replica $1$ and some in replica $2$. Imposing eq.~\eqref{solution_delta_U1_3d} as a solution of the constraint in this geometry, we obtain the following equations
\begin{align*}
k^{(1)}_{i,0} = \theta^{(2)}_{j+\hat{0},1} + \theta^{(2)}_{j+\hat{0}+\hat{1},2} - \theta^{(2)}_{j+\hat{0}+\hat{2},1} - \theta^{(2)}_{j+\hat{0},2}, \quad k^{(1)}_{i-\hat{0},0} = \theta^{(1)}_{j,2} + \theta^{(1)}_{j+\hat{2},1} - \theta^{(1)}_{j+\hat{1},2} - \theta^{(1)}_{j,1},\\
k^{(1)}_{i,1} = \theta^{(1)}_{j + \hat{1},2} + \theta^{(1)}_{j+\hat{1} + \hat{2},0} - \theta^{(2)}_{j+\hat{1}+\hat{0},2} - \theta^{(1)}_{j+\hat{1},0}, \quad k^{(1)}_{i-\hat{1},1} = \theta^{(1)}_{j,0} + \theta^{(1)}_{j+\hat{0},2} - \theta^{(1)}_{j+\hat{2},0} - \theta^{(1)}_{j,2},\\
k^{(1)}_{i,2} = -\theta^{(1)}_{j + \hat{2},1} - \theta^{(1)}_{j+\hat{2} + \hat{1},0} + \theta^{(2)}_{j+\hat{2}+\hat{0},1} + \theta^{(1)}_{j+\hat{2},0}, \quad k^{(1)}_{i-\hat{2},2} = \theta^{(1)}_{j,1} + \theta^{(1)}_{j+\hat{1},0} - \theta^{(2)}_{j+\hat{0},1} - \theta^{(1)}_{j,0}.
\end{align*}
Using these equations, the constraint is identically satisfied if and only if
\begin{align}
\theta^{(1)}_{j+\hat{0},2} = \theta^{(2)}_{j+\hat{0},2}
\end{align}
(and similar relations for other replicas), i.e., if the gauge variables lying on the entangling surface are the same in the different replicas. Also in this case we obtain the same picture as in the $\Z_2$ gauge theory: gauge fields on the conical singularity belongs to all replicas, and they belong to $4n$ different plaquettes.

The construction is the same for the case of $\Z_N$ gauge theories, too. Also, it is easy to see that, one can also recover the central-plaquette geometry for $\Z_N$ and $\U(1)$ gauge theories starting from a spin model in which some spins lie on the conical singularity.

We conclude that, in all of the Abelian gauge theories we analyzed, the geometry of the replica lattice is the same, regardless of the details about the degrees of freedom. In particular, it is found that gauge fields on the entangling surface belong to a larger number of plaquettes.

\subsection{Higher dimensions and non-Abelian theories}
\label{subsec:higher_dimensions_and_non-Abelian_theories}

We conclude this section with some comments on generalizations of the previous analysis to other theories.

The generalization to higher dimensions is straightforward. In spacetime dimension $D$ the entangling surface is a codimension-$2$ manifold. Any $s$-simplex lying on the entangling surface has a larger coordination number, i.e., a larger number of $(s+1)$-simplices touching it. For example, a site on the conical singularity is connected to $(4n + 2D-4)$ sites: $2(D-2)$ on the entangling surface and $4n$ in the two transverse directions. A link on the entangling surface belongs to $(4n + 2D-6)$ different plaquettes, and so on.

The central-plaquette geometry can be generalized to higher dimensions, too. Consider an $s$-simplex with at its boundaries a set of $(s-1)$-simplices, with $s>1$. If the simplex encloses the conical singularity, the size of its boundary is enlarged, since it has to wind around the singularity $n$ times before closing on itself.

An important and less simple question is, whether it is possible to extend this construction to non-Abelian gauge theories. Duality transformations for non-Abelian gauge theories are known, but they are considerably more involved~\cite{Burgio:1999tg, Mathur:2005fb, Cherrington:2007ax, Cherrington:2007is, Cherrington:2008ey, Tagliacozzo:2014bta, Mathur:2015wba, Mathur:2021vbp}, and it is not straightforward how to apply them to derive a geometric construction for the dual replica space.

Note however that, from a purely geometric point of view, the key features that we discussed in this section for the construction of the dual replica space (namely, the enlarged number of nearest neighbors and the central-plaquette setup) can be directly implemented---and, in fact, have been implemented~\cite{Chen:2015kfa}---to study R\'enyi entropies in arbitrary gauge theories. Our analysis shows that, at least for Abelian gauge theories, such geometries are justified by the duality construction. While we have not proven that these geometries are the correct ones also for non-Abelian theories, the observation that the geometry of the replica space appears not to be dependent on the gauge group suggests that this may be the case.

\section{Connection with the Hamiltonian formalism}
\label{sec:connection_with_the_Hamiltonian_formalism}

In section~\ref{sec:duality_transformation} we applied the duality transformation to derive the geometry of the replica space for Abelian gauge theories, working in the Lagrangian formalism. Our derivation followed the path depicted in figure~\ref{fig:diagram_of_this_work}, from the lower left corner to the upper right corner, passing through the upper left corner: one starts from a quantum lattice model in the Hamiltonian formalism, where there exists a well-defined notion of Hilbert space, and chooses a factorization $\mathcal{H} = \mathcal{H}_A\otimes\mathcal{H}_B$. For the spin models considered here, such as the Ising model, the factorization can be performed without ambiguities. At this stage one can discretize the path integral that appears in the replica trick, namely
\begin{align}
\Tr\rho_A^n = \frac{1}{Z^n}\prod_{r=1}^n\sum_{\{\sigma^A_r\sigma^B_r\}} \mel{\sigma^A_r\sigma^B_r}{\exp(-\frac{H_r}{T})}{\sigma^A_{r+1}\sigma^B_r},
\label{discretization_of_replica_trick}
\end{align}
where $r$ labels the different replicas and $\left\{\ket{\sigma^A_r\sigma^B_r}\right\}$ denotes a complete set of states (where we made the bipartition of the system explicit). In the simple case of the $(1+1)$-dimensional Ising model, the Hamiltonian in each replica can be written as the sum of three terms, $H_A$, $H_B$, and $h$: the first two respectively describe the interactions between degrees of freedom inside $A$ only and in $B$ only, while $h$ represents the interactions between the degrees of freedom in $A$ and those in $B$: 
\begin{align}
H = H_A + H_B + h = \sum_{\langle i,j \rangle\in A} \sigma_i^z\sigma_j^z + g\sum_{i\in A}\sigma_i^x + \sum_{\langle i,j \rangle\in B} \sigma_i^z\sigma_j^z + g\sum_{i\in B}\sigma_i^x + \sum_{\substack{\langle a,b\rangle,\\a\in A, \, b\in B}}\sigma_a^z\sigma_b^z.
\end{align}
$\sigma^z$ and $\sigma^x$ are Pauli matrices and, in the boundary term $\sigma_a^z\sigma_b^z$, $a$ and $b$ are nearest neighbors. From eq.~\eqref{discretization_of_replica_trick} one can obtain the partition function of a statistical-mechanical system by dividing the (compact) time direction in $\Ntau$ intervals of length $a_\tau = 1/(T \Ntau)$, Trotterizing the exponential and inserting a resolution of the identity at each discrete time step. If we take, for definiteness, the $\ket{\sigma^A_r\sigma^B_r}$ to be eigenstates of $\sigma^z$, the hopping term $\sigma^z_i\sigma_j^z$ leads to a nearest-neighbor interaction in the spatial direction, while the $\sigma^x$ operators generate interactions in the temporal direction. The replica structure shown in figure~\ref{fig:replica_space_cut} emerges from terms such as
\begin{align}
\mel{\sigma^A_r\sigma^B_r;\tau = 1/T}{\exp(g\sigma^x)}{\sigma^A_{r+1}\sigma^B_r;\tau = 0},
\end{align}
where $\ket{\sigma^A_r\sigma^B_r;\tau}$ is a state inserted at Euclidean time $\tau$. The previous term leads to an interaction in the time direction between two spins in the same replica, if they belong to $B$, and between two different replicas, if they belong to $A$. From the discrete replica space one can then obtain the dual geometry as described in section~\ref{sec:duality_transformation}.

In this section we discuss how one can derive the same geometric picture by reversing the order of the steps, i.e., by first performing the duality transformation in the Hamiltonian description, and then discretizing the resulting path integral. For simplicity from now on we will discuss the case of the $(1+1)$-dimensional Ising model, but the discussion is easily generalizable to other spin models and higher dimensions.

For this purpose, instead of working with states in a Hilbert space, it is more convenient to work with algebras acting on that Hilbert space~\cite{Casini:2013rba}. Given a Hilbert space $\mathcal{H}$ and an algebra of observables $\mathcal{A}$, a bipartition of the system is now defined by identifying a subalgebra $\mathcal{A}_A$ of operators acting on the subsystem $A$. The reduced density matrix is then defined as the unique operator $\rho_A\in\mathcal{A}_A$ satisfying
\begin{align}
\Tr(\rho_A \mathcal{O}) = \expval{\mathcal{O}},
\end{align}
for all $\mathcal{O}\in\mathcal{A}_A$. In general $\rho_A$ can be written as a linear combination of the generators of the algebra. For the $(1+1)$-dimensional Ising model on a chain of $N$ sites, the maximal algebra one can define is generated by the two Pauli operators $\sigma^x$ and $\sigma^z$ on each site
\begin{align}
\mathcal{A} = \{\sigma^x_1, \dots ,\sigma^x_N, \sigma^z_1, \dots,\sigma^z_N\}.
\end{align}
The choice of a subalgebra which leads to eq.~\eqref{discretization_of_replica_trick} is the following: let $A$ be the set of sites with $1 \leq i \leq M < N$, then
\begin{align}
\mathcal{A}_A = \{\sigma^x_1, \dots ,\sigma^x_M,\sigma^z_1, \dots ,\sigma^z_M\},
\label{maximal_subalgebra_spin_chain}
\end{align}
which is the maximal subalgebra we can define on the subchain from site $1$ to site $M$. One can then act on the subalgebra defined in eq.~\eqref{maximal_subalgebra_spin_chain} with the Kramers--Wannier duality~\cite{Radicevic:2016tlt}: the subchain of $M$ sites maps to a subchain of $(M+1)$ sites whose edges lie on the entangling surface, and the algebra $\mathcal{A}_A$ maps to the non-maximal algebra (in this section we label dual sites with half-integers)
\begin{align}
\mathcal{A}_A^\star = \{\sigma^x_{3/2}, \dots ,\sigma^x_{M-1/2},\sigma^z_{1/2},\dots ,\sigma^z_{M+1/2}\},
\label{dual_algebra_without_edge_sigma_x}
\end{align}
where the $\sigma^x$ generators at the edges are removed. As a consequence, this algebra has now a center, i.e., a set of generators commuting with all the other generators of the algebra. Any operator belonging to an algebra with a center must be block-diagonal in a basis that diagonalizes the center, therefore the Hilbert space splits into superselection sectors, labeled by the eigenvalues of the generators belonging to the center. Therefore, states with reduced density matrices in this algebra must be eigenstates of the generators of the center, in this case $\sigma^z_{1/2}$ and $\sigma^z_{M+1/2}$.

The path-integral representation of $\Tr\rho_A^n$ in the presence of superselection sectors was analyzed in ref.~\cite{Lin:2018bud}. The key difference with respect to eq.~\eqref{discretization_of_replica_trick} is that one has to keep different superselection sectors separated in $\Tr\rho_A^n$. One can first compute the path integral in a fixed sector, then sum the result over different sectors. Importantly, the sector is the same for all replicas. For the algebra defined in eq.~\eqref{dual_algebra_without_edge_sigma_x}, this leads to
\begin{align}
\Tr\rho_A^n = \frac{1}{Z^n}\sum_{s,s^\prime}\prod_{r=1}^n\sum_{\sigma^A_r\sigma^B_r}\mel{\sigma^A_r \sigma^B_r}{\exp\left(-\frac{H_r}{T}\right)}{\sigma^A_{r+1}\sigma^B_r}^{(s,s^\prime)}.
\end{align}
In this expression the eigenvalues of $\sigma^z_{1/2}$ and $\sigma^z_{M+1/2}$ in the summand are fixed to be $s$ and $s^\prime$ respectively, and they are the same for all replicas. The $h$ term in the Hamiltonian, describing interactions between pairs of degrees of freedom of which one is in $A$, the other in $B$, can be written as
\begin{align}
h = g (\sigma^x_{1/2} + \sigma^x_{M+1/2}).
\end{align}
The difference with respect to the previous case is in the presence of terms such as
\begin{align}
\mel{\sigma^A_r\sigma^B_r;\tau = 0}{\exp\left(g\sigma^x_{1/2}\right)}{\sigma^A_r \sigma^B_r;\tau = a_{\tau}}^{(s,s^\prime)} = \exp\left(\sigma^{(r)}_{\tau=0}\sigma^{(r)}_{\tau=a_\tau}\right) \delta_{s,\sigma^{(r)}_{\tau=0}},
\end{align}
that is, for all replicas spins at sites $i=\frac{1}{2}$ and $i=M+\frac{1}{2}$ for $\tau = 0$ have the same value in all configurations that contribute to the partition function. Note that this happens only for $\tau = 0$, since at all other times resolutions of the identity have been inserted, leading to independent sums over all eigenvalues of $\sigma^z$ on each site.

Given that the spins on the entangling surface have the same value in all configurations, they are effectively the same spin: this is how a geometry including two spins with an enlarged number of nearest neighbors emerges directly in the Hamiltonian formalism.

The derivation of the replica space in the presence of superselection sectors discussed in this section is completely general, and can be applied to other spin systems, as well as to higher-dimensional theories. In higher dimensions the explicit derivation of the discretized path integral starting from the Hamiltonian of the theory is less straightforward, and we do not address the calculation here. However, it would be interesting to study how different choices of algebras for gauge theories map to different replica spaces. In particular, the main choices one can make are two: the electric-center algebra and the magnetic-center algebra~\cite{Casini:2013rba}; it seems natural to expect that these two choices correspond to the two different geometries we found in section~\ref{sec:duality_transformation}, namely the central-plaquette geometry and the one with the enlarged number of nearest neighbors, while combinations of the two choices of the center might map into combinations of the previous two geometries.

As a final comment, we stress that an algebraic choice in the Hamiltonian formalism is mapped to a purely geometric one in the Lagrangian formalism. In particular, different choices of the algebra $\mathcal{A}_A$ correspond to different geometries of the entangling surface. In both pictures, the differences between the various choices seem to be related to the ultraviolet physics of the model, and this is consistent with the expectation that in the continuum limit the universal part of the entanglement entropy, as well as properly regularized physical quantities such as the mutual information, are independent from the choice of the algebra.

\section{Monte Carlo results for the $\Z_2$ gauge theory in three dimensions}
\label{sec:monte_carlo}

In this section we present the results of a Monte Carlo study of the entropic c-function of the ground state of the three-dimensional $\Z_2$ gauge theory. An early study of the entanglement entropy in this model by means of Monte Carlo simulations was reported in ref.~\cite{Buividovich:2008gq}, while more recently this problem has been addressed in various studies based on tensor networks~\cite{Xu:2023zsz, Feldman:2024qif, Knaute:2024wfh} and exact diagonalization~\cite{Mueller:2021gxd}.

As discussed in refs.~\cite{Calabrese:2004eu, Calabrese:2009qy}, the replica trick allows one to express R\'enyi entropies in terms of ratios of partition functions, which are quantities that today can be computed to high precision by means of Monte Carlo simulations on the lattice. This is true also for the associated entropic c-functions, which can be written as
\begin{align}
C_n(l) = \frac{l^{D-1}}{|\partial A|}\frac{1}{n-1}\lim_{\epsilon\rightarrow 0}\frac{1}{\epsilon}\ln\frac{Z_n(l)}{Z_n(l+\epsilon)}.
\label{entropic_c-function_as_a_ratio_of_partition_functions}
\end{align}
In this work we make use of a highly efficient algorithm, based on Jarzynski's theorem~\cite{Jarzynski:1996oqb, Jarzynski:1997ef}, that we introduced in ref.~\cite{Bulgarelli:2023ofi}, generalizing previous work done in ref.~\cite{Alba:2016bcp}. Jarzynski's theorem is an exact equality between ratios of partition functions defined by equilibrium Boltzmann distributions and averages over non-equilibrium processes. Specifically, consider two probability distributions $p_0 = \exp(-S_0)/Z_0$ and $p_1 = \exp(-S_1)/Z_1$ which can be continuously connected by tuning a parameter $\lambda$, such that $S(\lambda = 0) = S_0$ and $S(\lambda=1) = S_1$; Jarzynski's theorem then states that 
\begin{align}
\frac{Z_1}{Z_0} = \left\langle e^{-W_{0\rightarrow 1}} \right\rangle
\end{align}
where $W_{0\rightarrow 1}$ is the (generalized, dimensionless) work performed to drive a system, initially at equilibrium, from $\lambda=0$ to $\lambda=1$, and the average is performed over an ensemble of different trajectories that the system follows in its out-of-equilibrium evolution. It is important to stress that the equality holds regardless of the specific details of how the system is driven off equilibrium: the equality holds irrespectively of the way in which $\lambda$ is varied, as long as this evolution is the same for all trajectories, and that the system is genuinely out of equilibrium for every $\lambda>0$. In recent years, this theorem has been harnessed in high-precision lattice simulations, to study a variety of different quantities~\cite{Caselle:2016wsw, Caselle:2018kap, Francesconi:2020fgi, Caselle:2022acb, Caselle:2023uel, Bonanno:2024udh}, including in computations of R\'enyi entropies through quantum Monte Carlo simulations~\cite{DEmidio:2019usm, Zhao:2021njg, Zhao:2021ghz, DaLiao:2023pdn, Pan:2023ysg, Deng:2023akc, Song:2023wlg, DEmidio:2024ggo}.

To obtain high-precision results for the $\Z_2$ gauge theory we exploit its duality to the three-dimensional Ising model---using, in particular, the fact that the entropic c-function computed in the two theories is the same, as shown by eq.~\eqref{entropic_c-function_under_duality_3d}. The geometry we simulated is a slab geometry, with the entangling surface lying along the links of the dual lattice.

A major advantage of directly simulating the Ising model is the possibility to use cluster algorithms, which essentially allow one to perform simulations arbitrarily close to critical points, without suffering from critical-slowing-down problems. In our simulations we used the Swendsen--Wang cluster algorithm~\cite{Swendsen:1987ce} in a parallelized version~\cite{Kalentev:2010ccl, Komura:2012gbs, Komura:2014cpf} adapted for GPU-accelerated supercomputers.

The non-equilibrium protocol we used to calculate the entropic c-function is the same as the one we introduced in ref.~\cite{Bulgarelli:2023ofi}. First, note that the ratio of partition functions appearing in eq.~\eqref{entropic_c-function_as_a_ratio_of_partition_functions} can be approximated on the lattice in terms of $Z_n(l)/Z_n(l+a)$, with $a$ denoting the lattice spacing (i.e., the shortest length that can be resolved on the lattice). We can then design a non-equilibrium evolution connecting the two probability distributions by varying the couplings at the edge of the entangling surface: starting from a slab of length $l+a$, we reduce the couplings of spins in replica $r$ near the entangling surface with spins in replica $r+1$ and, at the same time, we increase the coupling between spins in the same replica, decreasing the length of the slab by one lattice spacing.

We performed simulations in a geometry of two replicas to compute the entropic c-function associated to the second R\'enyi entropy. Simulating different values of $\beta$ and for various values of the lattice volume allowed us to perform both thermodynamic and continuum extrapolations. To the best of our knowledge, this is the first study in which such extrapolations are performed for a $(2+1)$-dimensional gauge theory. For the scale setting we followed ref.~\cite{Caselle:1995wn}, where the dependence of the critical value of the Ising coupling $\betac$ on the lattice size in the Euclidean-time direction in units of the lattice spacing, denoted as $\Ntau$, was determined with high precision for a wide range of values. The critical deconfining temperature of the dual gauge theory can then be obtained as $a\Tc = 1/N_{\tau,\mbox{\tiny{c}}}$, so that one can then express all of the other dimensionful quantities in the appropriate units of $\Tc$. To make sure that the system is not affected by undesired finite-temperature corrections, we performed our simulations on lattices whose extent in the Euclidean-time direction is at least ten times larger than the critical one. Further details on the scale setting and on the different extrapolations can be found in section~\ref{app:simulation_details} of the appendix.

Figure~\ref{fig:final_data_with_fit} shows a summary of our final results. The plotted data correspond to the function
\begin{align}
\bar{C}_2(x) \equiv \frac{C_2(x)}{C_2^{\text{CFT}}}, \quad C_2^{\text{CFT}} = C_2(0),
\end{align}
where $C_2^{\text{CFT}}$ was estimated in ref.~\cite{Kulchytskyy:2019hft}. The length of the cut is measured in units of two different mass scales, namely, the critical temperature $\Tc$ and the mass $\mg$ of the lightest glueball-like state in the theory, taken from ref.~\cite{Agostini:1996xy}.

The behavior of the entropic c-function of the ground state is known only for few strongly interacting theories. Some predictions for gauge theories can be derived from the holographic correspondence, using the Ryu--Takayanagi formula~\cite{Ryu:2006bv, Ryu:2006ef} to compute the entanglement entropy in terms of the area of minimal surfaces. The authors of ref.~\cite{Klebanov:2007ws} calculated the entanglement entropy for confining backgrounds in non-Abelian gauge theories similar to quantum chromodynamics (QCD), and in a slab geometry at different length scales, finding the following picture: at small length scales (where, due to asymptotic freedom, the theory is invariant under conformal symmetry), the entropic c-function is constant and proportional to the number of degrees of freedom---which, for $\SU(N)$ gauge theories in the large-$N$ limit, scales like the square of the number of color charges $N$. As the length of the slab is increased, the theory eventually hadronizes, and, since the number of hadronic degrees of freedom is independent from $N$, the entropic c-function is then expected to exhibit a sharp transition, from a regime in which its parametric dependence on $N$ is $\order{N^2}$ to one in which it scales as $\order{N^0}$. The length scale at which this transition occurs is controlled by a mass scale associated to the boundary theory. For the QCD-like theories considered in ref.~\cite{Klebanov:2007ws}, all of the typical energy scales are parametrically equivalent, and so is the scale associated to the transition of the entanglement entropy.

\begin{figure}[t]
\centering
\includegraphics[width=.8\textwidth]{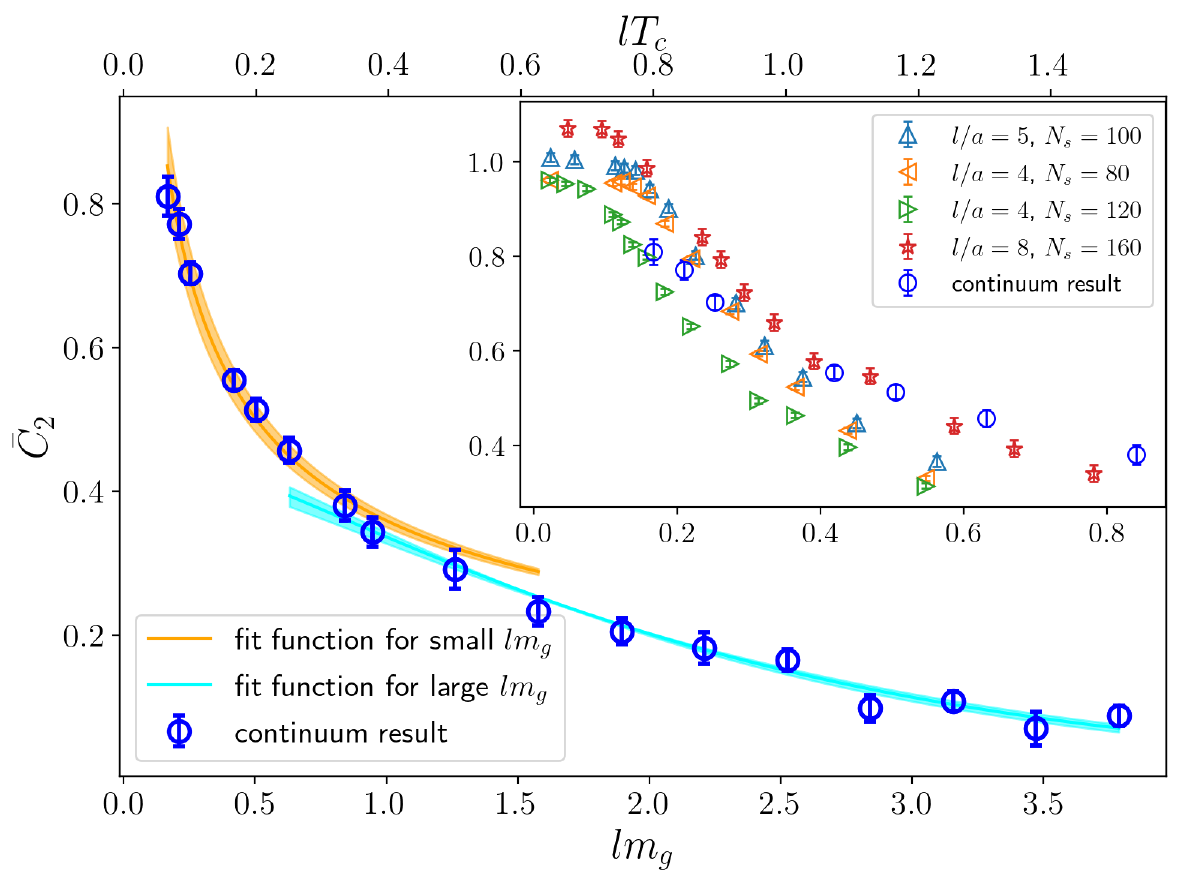}
\caption{Monte Carlo results for the entropic c-function of the second R\'enyi entropy. Inset: entropic c-function at small values of $l\mg$. Blue symbols denote the results extrapolated to the thermodynamic and continuum limits, while the other symbols correspond to results obtained from finite lattices.}
\label{fig:final_data_with_fit}
\end{figure}

In order to gain insight on the behavior of the entropic c-function in different regimes, we fitted it with different models. At large length scales, the theory can be modeled in terms of a gas of non-interacting glueballs~\cite{Mathieu:2008me, Caselle:2015tza}. On the other hand, for $(1+1)$-dimensional free scalar and Dirac fields of mass $m$ it was found that, for $n > 1$ and $lm \gg 1/(n-1)$, the entropic c-function admits the expansion~\cite{Casini:2005rm, Casini:2005zv}
\begin{align}
C_n^{(1+1)}(lm) = \exp(-2 m l) \left( 1 + \order{\frac{1}{ml}}\right).
\end{align}
In particular, for $n = 2$ the expansion is exact at leading order. Following ref.~\cite{Klebanov:2007ws}, one can think of a free scalar field in $2+1$ dimensions as a collection of infinite two-dimensional scalar fields with transverse momentum $\mathbf{k}$ and mass $m(\mathbf{k}) = \sqrt{m^2 + \mathbf{k}^2}$. Summing over the entropic c-functions of each mode leads to
\begin{equation}
C_n^{(2+1)}(lm) \sim l \int\dd \mathbf{k} \exp(-2\sqrt{m^2 + \mathbf{k}^2}l).
\end{equation}
Therefore we fitted our data with the \emph{Ansatz}
\begin{align}
\label{Ansatz}
f(x;A,\alpha) = A x \int \dd t \exp(-2\sqrt{1+t^2}\alpha x).
\end{align}
The function fits the data well from large values of $l\mg$ down to $l\mg \simeq 0.84$. The results of the fit are $A = 0.33(3)$, $\alpha = 0.360(19)$, with $\redchisq=0.82$ (where $\redchisq$ denotes the $\chi^2$ divided by the number of degrees of freedom of the fit). Note that the parameter $\alpha$ appearing in eq.~\eqref{Ansatz} controls the exponential decay of the entropic c-function at large distances, and it reduces to $1$ for a free scalar field. For the $\Z_2$ gauge theory we obtain a value of $\alpha$ which is smaller than in the free-field case, albeit roughly of the same order. A discrepancy with respect to the free case is expected, since the system we are considering is interacting.

Consider now the opposite regime (small $l\mg$); the recent ref.~\cite{Florio:2023mzk} reported a study of the two-mode logarithmic negativity in the Schwinger model, which was studied as a measure of entanglement at different length scales. It was found that, at distances shorter than the inverse of the mass of the lightest state in the continuum, the decay of the two-mode negativity is polynomial. Motivated by this observation, we fitted our data for small $l\mg$ with a power-law function
\begin{align}
f(x;B,c) = \frac{B}{x^c}.
\end{align}
The function fits well our results from small values of $l\mg$ up to $l\mg \simeq 1.26$. The results of the fit are $B = 0.360(9)$, $c = 0.48(2)$, with $\redchisq = 1.02$.

Note that a deviation from the power-law decay is expected at sufficiently short length scales, since $C_2(0)$ is finite. At very short distances, however, the simulations become more expensive in terms of computing resources, since one should get very close to the critical point of the theory and very large lattices are necessary to extrapolate to infinite volume. Despite these difficulties, looking at the data we obtained from finite lattices and for different values of the lattice spacing, as shown in figure~\ref{fig:final_data_with_fit}, one observes clear convergence towards a finite value for $l\mg\rightarrow 0$, even though some data sets appear to tend to a value slightly different from $1$. This is likely a systematic uncertainty due to the finiteness of the lattice spacing and/or of the length of the slab, which become particularly relevant effects in the $l\mg\rightarrow 0$ limit.

\section{Conclusions}
\label{sec:conclusions}

In this work we discussed the replica space and R\'enyi entropies of Abelian gauge theories; in particular, we exploited duality transformations relating them to spin models, and discussed the geometric features of this mapping. The advantages of this approach are twofold: on the one hand, the duality transformation can be explicitly performed in a replica space, leading to a derivation of the replica geometry which is valid for different gauge theories. On the other hand, the duality mapping allows one to compute entanglement measures for strongly interacting gauge theories by simulating spin models, which can be efficiently studied with Monte Carlo methods.

In the first part of this work we analytically performed the duality transformation for two- and three-dimensional spin systems with Abelian global symmetry group. The picture emerging from our study is a purely geometric one, in the sense that it does not depend on the specific degrees of freedom of the theory one is considering. Fields lying on the conical singularity belong to all replicas and the local action gets contributions from interactions with fields in all replicas; in a spin model, this translates into an enlarged number of nearest neighbors for the spins on the conical singularity, whereas in a gauge theory the gauge field therein contributes to a larger number of plaquettes. Note that this geometry realizes in a natural way a number of features of the replica space of a gauge theory that have already been discussed in past works, such as the fact that gauge transformations on the entangling surface must be the same on all replicas~\cite{Soni:2016ogt, Jokela:2022fvh} or the presence of Wilson loops that appear to be open, if one looks at a single replica, but close in a different one~\cite{Buividovich:2008gq}. The existence of a larger number of neighbors in the dual formulation of the theory has also been discussed recently in refs.~\cite{Abe:2023uan, Morikawa:2024zyd}.

If the conical singularity is placed in the dual lattice, plaquettes encircling the entangling surface have to loop around it $n$ times before closing onto themselves. This is the central-plaquette geometry discussed in ref.~\cite{Chen:2015kfa}. In the present work we focused mostly on the first geometry, since the Ising model we simulated is dual to a gauge theory with gauge fields on the conical singularity, without central plaquettes. It would be interesting to investigate how the results change if one uses a different geometry, and to test whether the same continuum limit is recovered.

In order to make contact with the Hamiltonian formalism, in section~\ref{sec:connection_with_the_Hamiltonian_formalism} we derived the replica geometry for the $(1+1)$-dimensional Ising model in presence of superselection sectors, finding the same geometry of section~\ref{sec:duality_transformation}, with spins on the entangling surface having an enlarged number of nearest neighbors. The derivation is completely general and in principle it might be extended to gauge theories with magnetic- and electric-center algebra. 

In the second part of our work, we numerically exploited the duality transformation, simulating the three-dimensional Ising model to compute the entropic c-function associated with the second R\'enyi entropy of the $\Z_2$ gauge theory. Despite its simplicity, this theory displays many phenomena that also characterize $\SU(N)$ gauge theories, such as a mass gap and confinement. Combining the efficient algorithm~\cite{Bulgarelli:2023ofi} based on Jarzynski's theorem~\cite{Jarzynski:1996oqb, Jarzynski:1997ef} and GPU parallelization, we were able to extrapolate our results to the infinite-volume and continuum limits, and to compare our final results with predictions based on the holographic correspondence~\cite{Klebanov:2007ws}.

As expected by semi-empirical arguments, the entropic c-function decays exponentially for large lengths of the slabs. At large length scales, the ground state of the gauge theory can be described as a gas of weakly interacting bosons, and indeed the exponential decay seems to be parametrically controlled by the mass of the lightest glueball, up to order-one numerical factors.

At short length scales, the entropic c-function scales as a power of the length of the slab. The exponent of the power-law decay we obtained fitting our data is compatible with $\frac{1}{2}$ and it might be related to the scaling dimension of some operator in the conformal field theory describing the three-dimensional Ising model at criticality. Note, however that such exponent is likely to be connected in a non-trivial way to the scaling dimension of the twist fields~\cite{Calabrese:2004eu, Cardy:2007mb}. The transition from power-law to exponential scaling is observed to occur around a length scale controlled by the mass of the lightest glueball, that is the mass gap of the theory. This behavior resembles the transition that was found in many holographic models dual to a confining background~\cite{Ryu:2006ef, Klebanov:2007ws, Kol:2014nqa} and also in Monte Carlo studies of $\SU(N)$ gauge theories~\cite{Buividovich:2008gq, Buividovich:2008kq, Buividovich:2008yv, Itou:2015cyu, Rabenstein:2018bri}. Moreover, we note that similar results were recently obtained from a study of the two-mode entanglement negativity in the Schwinger model~\cite{Florio:2023mzk}. 

In this work we studied the simplest, non-trivial gauge theory which admits a dual spin description. A natural generalization is to study the entropic c-function of other Abelian gauge theories by exploiting the duality transformation. In this respect, a particularly interesting theory is the $\U(1)$ gauge theory in $2+1$ dimensions. Moreover, as we previously stated, the same geometries we derived for Abelian theories can in principle be used to study R\'enyi entropies in non-Abelian gauge theories. We leave these directions of research for future work.

\subsection*{Acknowledgements}

We thank R.~Amorosso, M.~Caselle, I.~Klebanov, S.~K\"{u}hn, E.~Mazenc, A.~Nada, J.~Subils and L.~Tagliacozzo for useful comments and discussions. This work of has been partially supported by the Spoke 1 ``FutureHPC \& BigData'' of the Italian Research Center on High-Performance Computing, Big Data and Quantum Computing (ICSC) funded by MUR (M4C2-19) -- Next Generation EU (NGEU), by the Italian PRIN ``Progetti di Ricerca di Rilevante Interesse Nazionale -- Bando 2022'', prot. 2022TJFCYB, and by the ``Simons Collaboration on Confinement and QCD Strings'' funded by the Simons Foundation. The simulations were run on CINECA computers. We acknowledge support from the SFT Scientific Initiative of the Italian Nuclear Physics Institute (INFN).


\appendix

\section{Duality transformation and boundary conditions}
\label{app:topology}
\renewcommand{\theequation}{A.\arabic{equation}}
\setcounter{equation}{0}

\subsection*{Ising model in two dimensions}

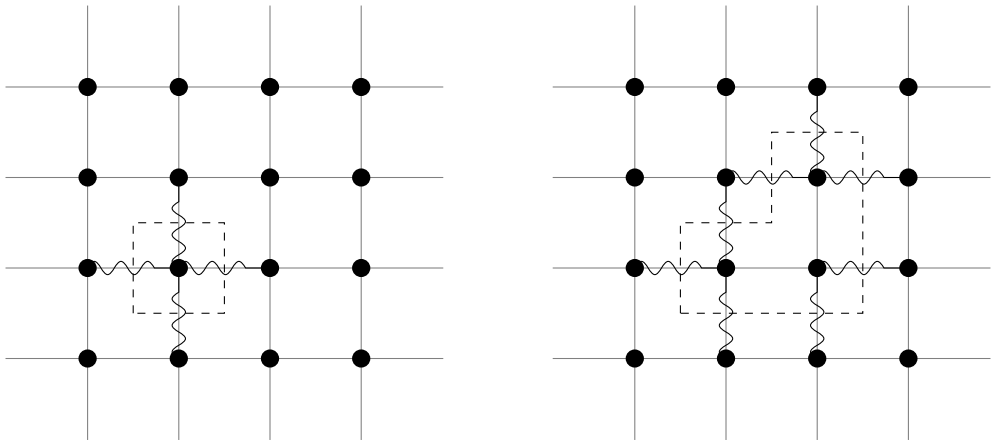
\begin{figure}[t]
\centering
\begin{tikzpicture}[scale=1.2, every node/.style={scale=1.2}]
\draw[step = 1cm, gray, thin] (-.9, -.9) grid (3.9, 3.9);
\foreach \x in {0,...,3}{
  \foreach \y in {0,...,3}{
    \fill[color=black] (\x, \y) circle (.1);
  }
}
\draw[step = 1cm, gray, thin] (5.1, -.9) grid (9.9, 3.9);
\foreach \x in {6,...,9}{
  \foreach \y in {0,...,3}{
    \fill[color=black] (\x, \y) circle (.1);
  }
}
\draw[thin, dashed] (.5,.5)--(.5,1.5)--(1.5,1.5)--(1.5,.5)--(.5,.5);
\draw[thin, dashed] (6.5,.5)--(6.5,1.5)--(7.5,1.5)--(7.5,2.5)--(8.5,2.5)--(8.5,.5)--(6.5,.5);
\draw[snake= coil, segment aspect = 0] (1,0) -- (1,1);
\draw[snake= coil, segment aspect = 0] (1,1) -- (1,2);
\draw[snake= coil, segment aspect = 0] (0,1) -- (1,1);
\draw[snake= coil, segment aspect = 0] (1,1) -- (2,1);
\draw[snake= coil, segment aspect = 0] (7,0) -- (7,1);
\draw[snake= coil, segment aspect = 0] (6,1) -- (7,1);
\draw[snake= coil, segment aspect = 0] (7,1) -- (7,2);
\draw[snake= coil, segment aspect = 0] (8,0) -- (8,1);
\draw[snake= coil, segment aspect = 0] (8,1) -- (9,1);
\draw[snake= coil, segment aspect = 0] (8,2) -- (9,2);
\draw[snake= coil, segment aspect = 0] (8,2) -- (8,3);
\draw[snake= coil, segment aspect = 0] (7,2) -- (8,2);
\end{tikzpicture}
\caption{Frustrations in the direct lattice (wavy links) and the corresponding closed paths in the dual lattice (dashed lines).}
\label{fig:direct_frustrations_and_dual_loops}
\end{figure}

In order to discuss how different boundary conditions on a finite lattice affect the duality transformation, we first slightly generalize the $D=2$ Ising model, allowing for the presence of a set of antiferromagnetic couplings; this can be done by introducing a background $\Z_2$ gauge field associated to the links of the lattice
\begin{align}
Z(\beta;\{\tau\}) = \sum_{\{\sigma\}}\exp\left(\beta\sum_{i,\mu}\sigma_i \tau_{i,\mu}\sigma_{i+\hat{\mu}}\right),
\label{partition_function_Ising_with_background_gauge_2d}
\end{align}
which reduces to the partition function of the Ising model when $\tau_{i,\mu} = 1$ for all $i$ and for all $\mu$. In addition to the global $\Z_2$ symmetry, the model defined by eq.~\eqref{partition_function_Ising_with_background_gauge_2d} has a local $\Z_2$ symmetry: changing the sign of the gauge fields defined on all links touching an arbitrary site, the partition function of the model is left invariant. For future convenience, note that a generic set of frustrations that is gauge-equivalent to a configuration without any frustrations at all is dual to a set of closed loops in the dual lattice, where the loops are made of links piercing the frustrated links in the direct lattice (see figure~\ref{fig:direct_frustrations_and_dual_loops}).

The duality transformation can then be performed in the same way as in section~\ref{sec:duality_transformation}, absorbing the gauge field, which is non-dynamical, in the definition of $C_k(\beta \tau)$. In this way the (local) dual coupling depends on the gauge field as
\begin{align}
(\beta \tau)^\star = -\frac{1}{2}\ln\tanh(\beta \tau) = \beta^\star -\frac{i\pi}{2}\frac{\tau-1}{2} = \beta^\star - \frac{i\pi}{2}\tau^\star, \qquad \mbox{with $\tau^\star\in\{0,1\}$}.
\label{dual_background_gauge_coupling}
\end{align}
Consider now a lattice with periodic boundary conditions in all directions. Using the fact that
\begin{align}
\exp(-\frac{i\pi}{2}s_i s_{i+\hat{\mu}}) = -is_i s_{i+\hat{\mu}},
\label{exp_pi_s_equal_minus_s}
\end{align}
the partition function can be written in terms of the dual variables as
\begin{align}
Z(\beta; \{\tau\}) = \frac{1}{2}(\sinh 2\beta^\star)^{-\lam}\sum_{\{s\}}\prod_{i,\mu}(s_i s_{i+\hat{\mu}})^{\tau^\star_{i,\mu}} \exp(\beta^\star s_i s_{i+\hat{\mu}}).
\label{dual_eq_with_background_gauge_fields_2d}
\end{align}
Note that for $\tau^\star = 0$ one recovers eq.~\eqref{dual_Ising_2d}. Let us now inspect the term $(s_i s_{i,\mu})^{\tau^\star_{i,\mu}}$: if the direct link is frustrated, the dual variables orthogonal to that link appear in the dual partition function as operator insertions. Consider a model with a set of frustrations that is gauge-equivalent to a configuration with no frustrations at all; as previously noted, such sets are dual to closed paths, therefore all operator insertions appear twice in eq.~\eqref{dual_eq_with_background_gauge_fields_2d} and one obtains again eq.~\eqref{dual_Ising_2d}.

It is now clear that a set of frustrations that is inequivalent to $Z(\beta)$ is dual to a set of open paths in the dual lattice. This leads to the well-known result
\begin{align}
\frac{Z(\beta; \{\tau\})}{Z(\beta)} = \left\langle \prod_{i,\mu} (s_i s_{i+\hat{\mu}})^{\tau^\star_{i,\mu}} \right\rangle,
\end{align}
namely a generic spin correlator in the $D=2$ Ising model can be expressed in terms of the ratio of the partition function with frustrations piercing paths that connect the various insertions, over the partition function without frustrations.

From the invariance of the partition function under insertion of a closed loop of frustrations which is contractible to a point, it follows that the constraint appearing in eq.~\eqref{Kronecker_delta_mod_2} admits additional solutions, beside the one discussed in section~\ref{sec:duality_transformation}, if the topology of the system is such that there exist non-contractible closed loops. In the latter case, the constraint will have different solutions, related to the inequivalent partition functions with the insertions of topological lines of frustrations along non-contractible paths. In general, if we start from the Ising model without frustrations, the general duality transformation reads
\begin{align}
Z(\beta) = \frac{1}{2[\sinh (2\beta^\star)]^{\lam}} Z^\star(\beta^\star).
\end{align}
For example, in the geometry of a torus
\begin{align}
Z^\star(\beta^\star)\propto Z_{pp}(\beta^\star) + Z_{ap}(\beta^\star) + Z_{pa}(\beta^\star) + Z_{aa}(\beta^\star),
\label{partition_function_with_topological_sectors_2d}
\end{align}
where $Z_{ij}(\beta)$ is the partition function with periodic ($p$) or antiperiodic ($a$) boundary conditions in directions $i$ and $j$; antiperiodic boundary conditions are equivalent to the insertion of frustrations along a non-contractible path.\footnote{More generally, antiperiodic boundary conditions are equivalent to the insertion of frustrations along any odd number of non-contractible paths that can be deformed into each other, since the effects of the frustrations on pairs of paths that can be deformed into each other cancel. Similarly, periodic boundary conditions correspond to the insertion of frustrations along any even number of non-contractible paths that can be deformed into each other.}

Note that topology affects the calculation of R\'enyi entropies, because the contributions from the partition functions with the different types of boundary conditions have to be taken into account. However, in some regions of the parameters space the contribution from antiperiodic boundary conditions is suppressed. In particular, antiperiodic boundary conditions generate phase boundaries, which are exponentially suppressed in the ordered phase. Therefore, up to numerical constants
\begin{align}
Z(\beta < \betac)\simeq Z_{pp}^\star(\beta^\star > \betac^\star).
\end{align}
This fact can be seen also at the level of the R\'enyi entropies. Consider eq.~\eqref{Renyi_entropy_under_duality_2d}: in the limit $\beta\rightarrow 0$ it gives the expected result:
\begin{align}
S_n(\beta = 0) = 0, \qquad S^\star_n(\beta^\star\rightarrow\infty) = \ln 2,
\end{align}
indeed at $\beta = 0$ the ground state of the Ising model is in a product state, while for $\beta^\star \rightarrow \infty$ the entropy takes a contribution from the two degenerate vacua. However, in the opposite limit, we do not recover the same result, rather:
\begin{align}
S_n(\beta \rightarrow \infty) = \ln 2, \qquad S^\star_n(\beta^\star = 0) = \ln 4.
\end{align}
The $\ln 4$ term can be related to the existence of the four different topological sectors appearing in eq.~\eqref{partition_function_with_topological_sectors_2d}.

\subsection*{Ising model in three dimensions}

The previous discussion can be easily generalized to three dimensions by looking for other solutions of the Kronecker-delta constraint in the case of the $\Z_2$ gauge theory. With the same logic as before, consider the following gauge transformation of the $\Z_2$ gauge theory: given any link of the lattice, we frustrate all four plaquettes sharing that link; it is easy to see that such transformation is a symmetry of the partition function. Moreover, the set of frustrated plaquettes is pierced by a closed path in the dual lattice.

New solutions to the constraint imposed by the Kronecker delta appearing in eq.~\eqref{Kronecker_delta_mod_2} can be obtained by choosing any set of closed paths in the lattice dual to the one of the $\Z_2$ gauge theory and introducing antiferromagnetic couplings for all plaquettes pierced by those paths.

If the paths are all contractible, then the resulting partition function is equivalent to the partition function of the $\Z_2$ gauge theory, therefore in a lattice with trivial topology there are no further solutions. If the topology is non-trivial, on the other hand, the general solution of the Kronecker-delta constraint corresponds to the sum of partition functions with insertions of frustrations along all of the possible types of non-contractible paths. For example, in the geometry of a three-dimensional torus, the partition function dual to the Ising model reads
\begin{align}
Z^{\Z_2}(\beta^\star) \propto Z^{\Z_2}_{+++}(\beta^\star) + Z^{\Z_2}_{++-}(\beta^\star) + \dots + Z^{\Z_2}_{+--}(\beta^\star) + Z^{\Z_2}_{---}(\beta^\star),
\end{align}
where $Z^{\Z_2}_{ijk}$ is the partition function of the gauge theory with ($-$) or without ($+$) the insertion of a topological line of frustrations along a non-contractible path in each of the three directions.

The same derivation can be carried out if one starts from the $\Z_2$ gauge theory to derive the Ising model as the dual theory. Like in the two-dimensional case, the three-dimensional Ising model is invariant if one frustrates all links touching one site. The difference from the two-dimensional case is that in three dimensions the set of frustrations is dual to a closed surface. Different solutions to the Kronecker-delta constraint are therefore associated with the partition functions of systems with frustrations that are dual to a set of closed surfaces. As a consequence, for a lattice with non-trivial topology one has to sum over all possible partition functions with insertions of non-contractible surfaces. In the geometry of a three-dimensional torus, one then recovers the well-known result
\begin{align}
Z(\beta) \propto Z_{ppp}(\beta) + Z_{ppa}(\beta) + \dots + Z_{paa}(\beta) + Z_{aaa}(\beta).
\label{duality_partition_function_torus_3d}
\end{align}
As in the two-dimensional case, partition functions with antiperiodic boundary conditions are suppressed in the confining phase of the gauge theory ($\beta^\star < \betac^\star$), which corresponds to the ordered phase of the spin model ($\beta > \betac$). Therefore
\begin{align}
Z^{\Z_2}(\beta^\star < \betac^\star) \simeq Z_{ppp}(\beta>\betac).
\end{align}
This is the reason why, in our Monte Carlo study of the entropic c-function of the $\Z_2$ gauge theory, we neglected contributions from topologically non-trivial sectors. In addition, the previous arguments explain why in section~\ref{sec:duality_transformation} we found that the strong-coupling result expressed by eq.~\eqref{topological_EE_from_duality} is consistent with the prediction that can be derived in the Hamiltonian formalism, while the weak-coupling one is not. In the weak-coupling regime, the result is spoiled by the effects of non-trivial topology, which are no longer suppressed.

\section{Duality in two-dimensional $\Z_N$ and $\U(1)$ models}
\label{app:duality_ZN_U1_2d}
\renewcommand{\theequation}{B.\arabic{equation}}
\setcounter{equation}{0}

In this section we discuss the duality transformation in systems with $\Z_N$ and $\U(1)$ symmetry in a replica space, as a natural generalization of our analysis for the Ising model. A relevant question we want to address here is whether the duality on a replica manifold is affected by the nature of the degrees of freedom of the theory, or just depends on geometric aspects, namely the presence of branch-point singularities. Therefore, for these models we will not keep track of the numerical prefactors in the partition function and we will just focus on the geometry of the dual lattice.

We focus on a specific class of models, the so-called clock models, defined by the partition function
\begin{align}
Z = \sum_{\{q\}}\exp\left[ \beta\sum_{i,\mu}\cos(\frac{2\pi}{N}\Delta_\mu q_i) \right],
\label{partition_function_clock_model_2d}
\end{align}
where $q\in\{0,1,\dots,N-1\}$ and $\Delta_\mu q_i \equiv q_i - q_{i+\hat{\mu}}$. The Boltzmann factor can be Fourier-transformed as
\begin{align}
\exp\left[ \beta\cos(\frac{2\pi}{N}\Delta_\mu q_i) \right] = \sum_{k=0}^N C_k(\beta) \exp(i\frac{2\pi}{N}k\Delta_\mu q_i),
\end{align}
where the $C_k(\beta)$ are the coefficients of the expansion. The partition function then becomes
\begin{align}
Z &= \sum_{\{q\}} \prod_{i,\mu}\sum_{k_{i,\mu}}C_{k_{i,\mu}}(\beta)\exp\left( i\frac{2\pi}{N} k_{i,\mu}\Delta_\mu q_i \right) \\
&= \sum_{\{q\}}\sum_{\{k\}}\prod_{i,\mu} C_{k_{i,\mu}}(\beta)\exp\left( -i\frac{2\pi}{N} q_i \sum_{\mu}\Delta_\mu k_{i,\mu} \right) \\
&= \sum_{\{k\}}\prod_{i,\mu} C_{k_{i,\mu}}(\beta) N\delta_N\left( \sum_\mu \Delta_\mu k_{i,\mu} \right).
\label{constraint_for_ZN_2d}
\end{align}
In the second line we used the discretized version of integration by parts, while in the last expression we used the Fourier decomposition of the Kronecker delta enforcing the constraint $\Delta_\mu k_{i,\mu} = 0$~mod~$N$. Let us start by considering the simplest case of an infinite lattice without any replica structure.

It is easier to solve the constraint in terms of another set of variables $\bar{k}$, which are equal to $k$~mod~$N$ and satisfy $\Delta_\mu \bar{k}_{i,\mu} = 0$ directly. Since there is a unique change of variable from $k$ to $\bar{k}$ (up to a $\Z_N$ transformation), the representation is the same in terms of $\bar{k}$. Using the $\bar{k}$ variables, the constraint can be solved by introducing a set of integer-valued spins $\phi$ on the sites of the dual lattice and imposing
\begin{align}
\bar{k}_{i,\mu} = \epsilon_{\mu\nu}\Delta_\nu \phi.
\label{solution_of_the_constraint_for_Z_N}
\end{align}
From eq.~\eqref{solution_of_the_constraint_for_Z_N} one then obtains
\begin{align}
Z \propto \sum_{\{\phi\}}\prod_{i,\mu}\exp\left\{ \ln C_{\epsilon_{\mu\nu}\Delta_\nu\phi}(\beta) \right\}.
\end{align}
This expression can be manipulated further; for $N=2$, $3$, or $4$, the model is self-dual.

Consider now a system of $n$ replicas. It is clear that also in this case the duality is only affected by the presence of branch points, therefore we focus our attention on the spins close to one edge of the cut. The geometry is the same as in figure~\ref{fig:duality_close_to_the_cut_Ising_2d}, but with the variables $s$ replaced by $\phi$. We can solve the constraint at site $i$ imposing eq.~\eqref{solution_of_the_constraint_for_Z_N},
\begin{align*}
\bar{k}_{i,0}^{(1)} &= \phi_{j+\hat{0}}^{(2)} - \phi_{j+\hat{0}+\hat{1}}^{(2)}, \\
\bar{k}_{i,1}^{(1)} &= \phi_{j+\hat{0}+\hat{1}}^{(2)} - \phi_{j+\hat{1}}^{(1)}, \\
\bar{k}_{i-\hat{0},0}^{(1)} &= \phi_{j+\hat{1}}^{(1)} - \phi_j^{(1)}, \\
\bar{k}_{i-\hat{1},1}^{(1)} &= \phi_j^{(1)} - \phi_{j+\hat{0}}^{(1)},
\end{align*}
which satisfy $\Delta_\mu\bar{k}_{i^{(1)},\mu}=0$ only if $\phi_{j+\hat{0}}^{(1)}=\phi_{j+\hat{0}}^{(2)}$. We see again that the dual spin defined on the branch point is shared by all replicas and has a larger number of nearest neighbors. Note that this geometric picture does not depend on the specific choice of the symmetry group. 

The case of a model with $\U(1)$ symmetry is straightforward, since the partition function is equivalent to eq.~\eqref{partition_function_clock_model_2d} with continuous variables, and the duality transformation is performed following the same steps, leading again to a geometry where the spin on the branch point singularity has a larger number of nearest neighbors.

\section{Derivation of the model dual to $\U(1)$ gauge theory}
\label{app:derivation_dual_U1}
\renewcommand{\theequation}{C.\arabic{equation}}
\setcounter{equation}{0}

In this section we derive the expression of the partition function of the spin theory dual to the $\U(1)$ gauge theory in three dimensions, eq.~\eqref{dual_to_U1}. Starting from the partition function
\begin{align}
Z^{U(1)} = \int D\theta \exp\left(\beta\sum_{\Box}\cos\theta_\Box\right)
\label{ZU1}
\end{align}
and using the expansion of the Boltzmann factor in Fourier modes: 
\begin{align}
\exp(\beta\cos x) = \sum_{n=-\infty}^{\infty} I_n(\beta)\exp(inx),
\label{Fourier_transform_U1}
\end{align}
eq.~\eqref{ZU1} can be rewritten as
\begin{align}
Z^{U(1)} &= \int D\theta \prod_\Box \sum_{k_\Box} I_{k_\Box}(\beta) e^{ik_\Box\theta_\Box} = \sum_{\{k\}}\prod_\Box I_{k_\Box}(\beta) \int \dd\theta_{i,\mu}e^{i\theta_{i,\mu}\sum k_{\Box}} \propto \\
 & \propto \sum_{\{k\}} \prod_\Box I_{k_\Box}(\beta)\delta\left(\sum k_\Box \right).
\end{align}
In this expression, the $k_\Box$ variables are associated to the plaquettes, and the sum $\sum k_\Box$ is over the four plaquettes sharing one link. The constraint has therefore the same structure of the constraint in eq.~\eqref{constraint_for_ZN_2d}, and thus admits the same solution. By introducing integer-valued spins $\phi$ on the sites of the dual lattice, one can write
\begin{align}
k_\Box \equiv k_{\mu\nu} = \epsilon_{\mu\nu\lambda}\Delta_\lambda\phi,
\end{align}
that is, a plaquette variable $k$ is dual to a difference of spins defined on the ends of the dual link orthogonal to the plaquette. Therefore the partition function can be written as
\begin{align}
Z^{U(1)} \propto \sum_{\{\phi\}}\prod_{\mu\nu i}I_{\epsilon_{\mu\nu\lambda}\Delta_\lambda\phi_i}(\beta).
\end{align}

\section{Details of the simulations}
\label{app:simulation_details}
\renewcommand{\theequation}{D.\arabic{equation}}
\setcounter{equation}{0}

\begin{figure}[t]
\centering
\begin{subfigure}{.47\textwidth}
\includegraphics[width=\textwidth]{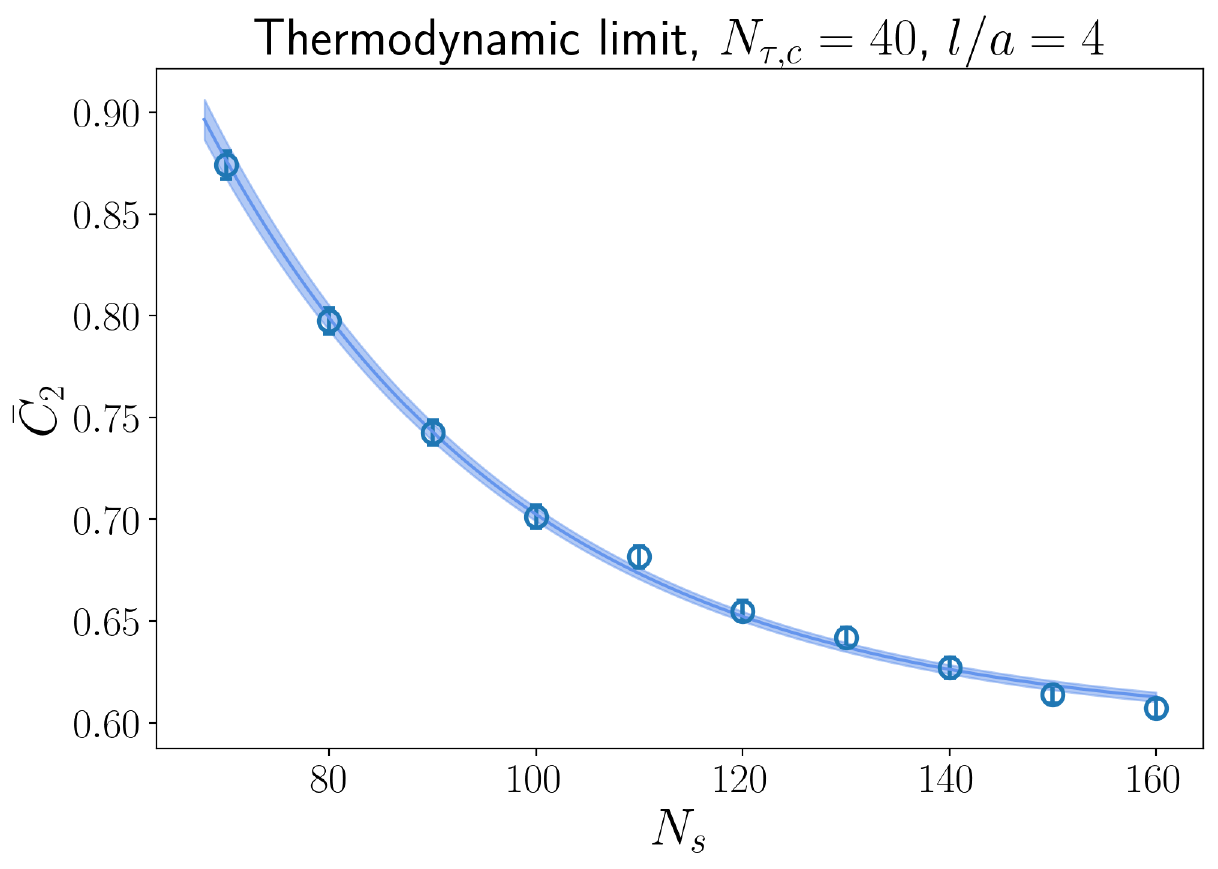}
\end{subfigure}
\begin{subfigure}{.47\textwidth}
\includegraphics[width=\textwidth]{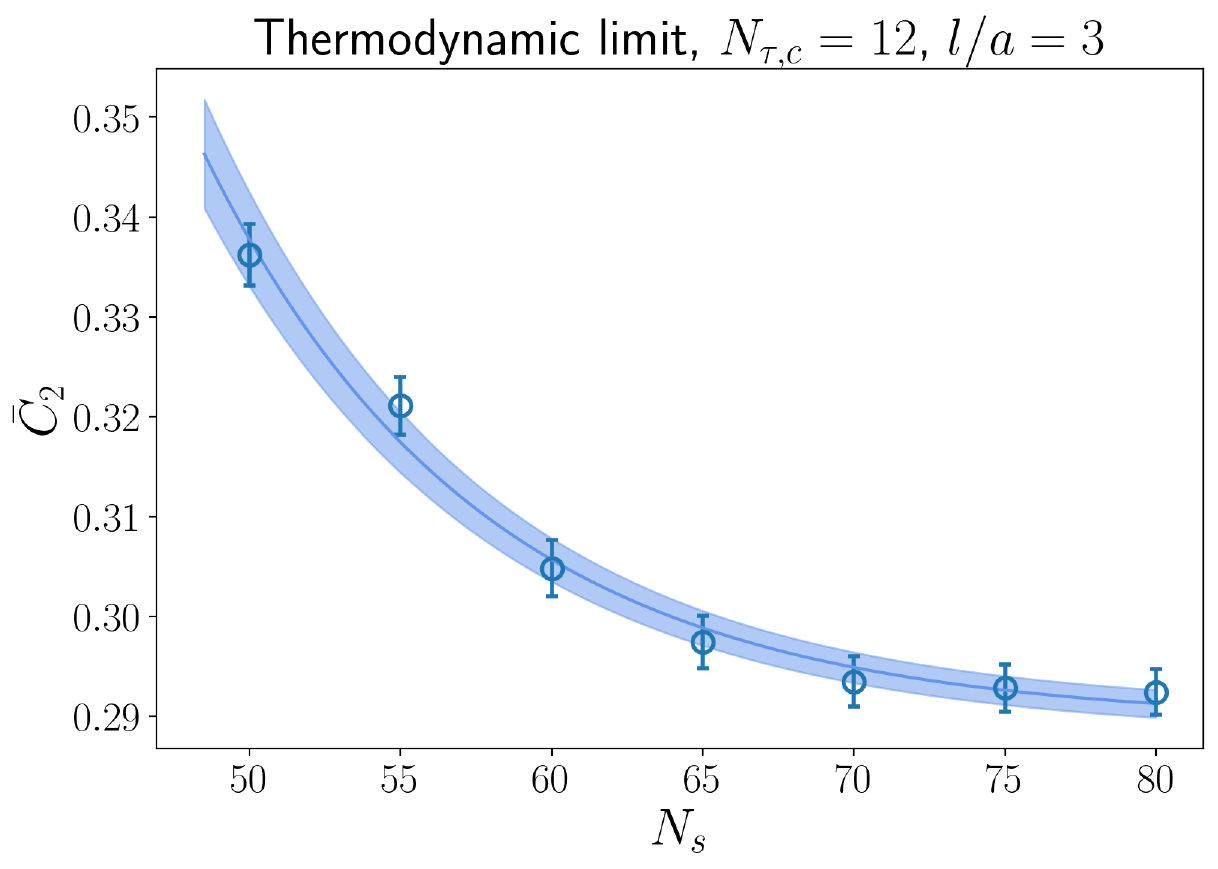}
\end{subfigure}
\begin{subfigure}{.47\textwidth}
\includegraphics[width=\textwidth]{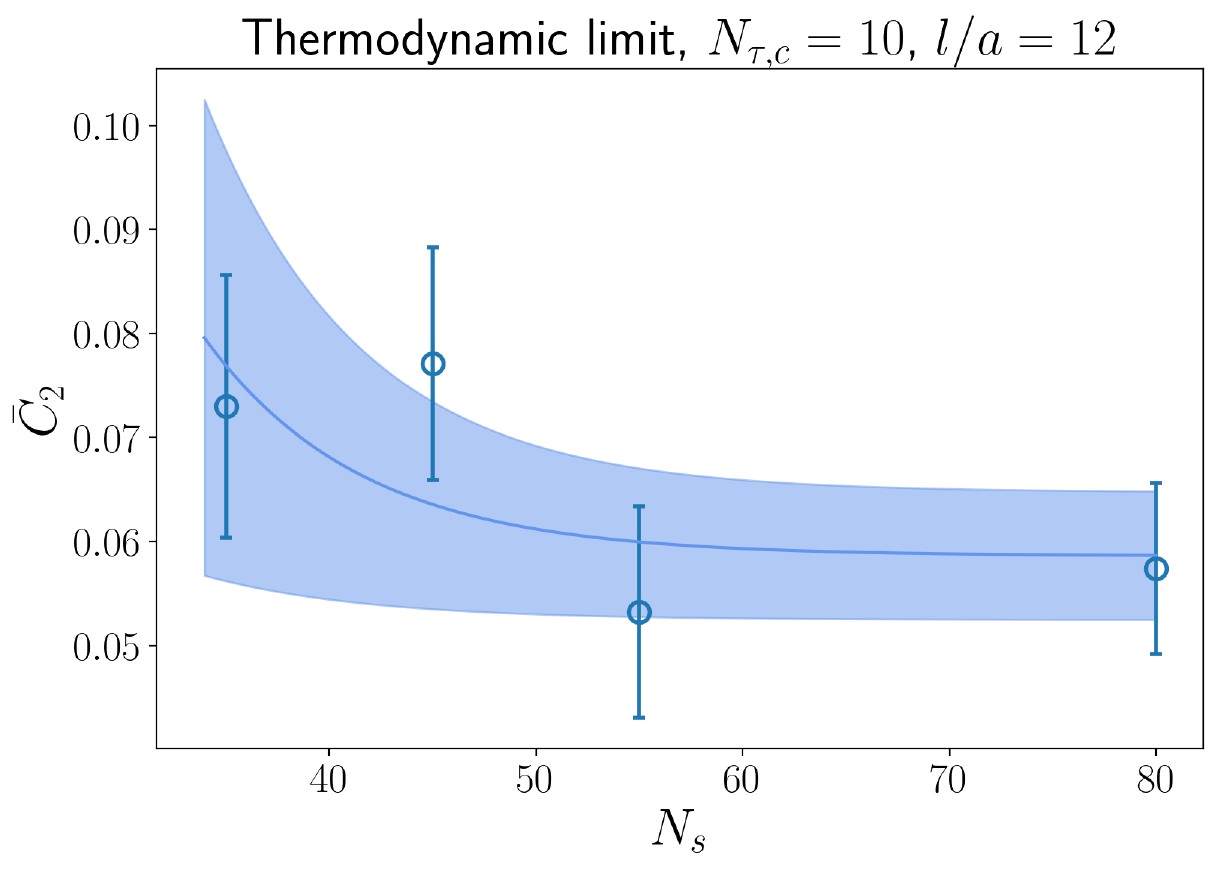}
\end{subfigure}
\caption{Thermodynamic limit for different values of the lattice spacing and length of the slab.}
\label{fig:thermodynamic_limit}
\end{figure}

In order to convert our results from lattice units to physical units, we set the scale using the study of the Ising model in three dimensions reported in ref.~\cite{Caselle:1995wn}, where the values of $\beta$ at which the deconfinement transition occurs were computed numerically, for a range of temporal extents up to $\Ntau = 16$. To perform simulations even closer to the critical point and access finer lattices, we extrapolated the results of ref.~\cite{Caselle:1995wn}, using eq.~(22) in the same paper, to larger $\Ntau$ values, as shown in table~\ref{tab:critical_values}. Note that here and in the following we are implicitly trading the $a \to 0$ limit for the equivalent $\Ntau \to \infty$ limit. For additional information on our simulation, see table~\ref{tab:thermodynamic_limit}.

At fixed values of the lattice spacing and for fixed length of the slab we performed an infinite-spatial-volume extrapolation, keeping $\Ntau$ fixed to at least $10N_{\tau,\mbox{\tiny{c}}}$ and $\Ntau$ larger than the spatial extent (in units of the lattice spacing) $N_s$. For values of $\beta$ close to the critical point, finite-volume effects are non-negligible, and an extrapolation is necessary, whereas for larger values of $(\beta - \betac)$ we were able to simulate physical volumes that are compatible with the infinite-volume limit within their uncertainties. This allowed us to bypass the need of extrapolation to the thermodynamic limit for simulations of large slabs, where the signal is poor and large statistics is required.

The fitting function we used for the infinite-volume extrapolation, shown in figure~\ref{fig:thermodynamic_limit}, is
\begin{align}
f(L; c, A, M) = c + A \exp(-ML),
\label{fit_function_for_thrmodynamic_extrapolation}
\end{align} 
where $L = aN_s$. The fit parameter $c$ is the infinite-volume result, while $M$ is a mass parameter which we imposed to be the same for all the extrapolations, by performing a global fit. The value we obtained is $M / \Tc = 1.31(2)$ with a global $\redchisq=1.87$. The fit function in eq.~\eqref{fit_function_for_thrmodynamic_extrapolation} is empirical, since no study on the scaling of the entropic c-function at finite volumes exists in literature. A better understanding of the scaling properties of such non-local quantities would be important, to have better control on the extrapolations from results at smaller volumes.

\begin{figure}[t]
\centering
\begin{subfigure}{.47\textwidth}
\includegraphics[width=\textwidth]{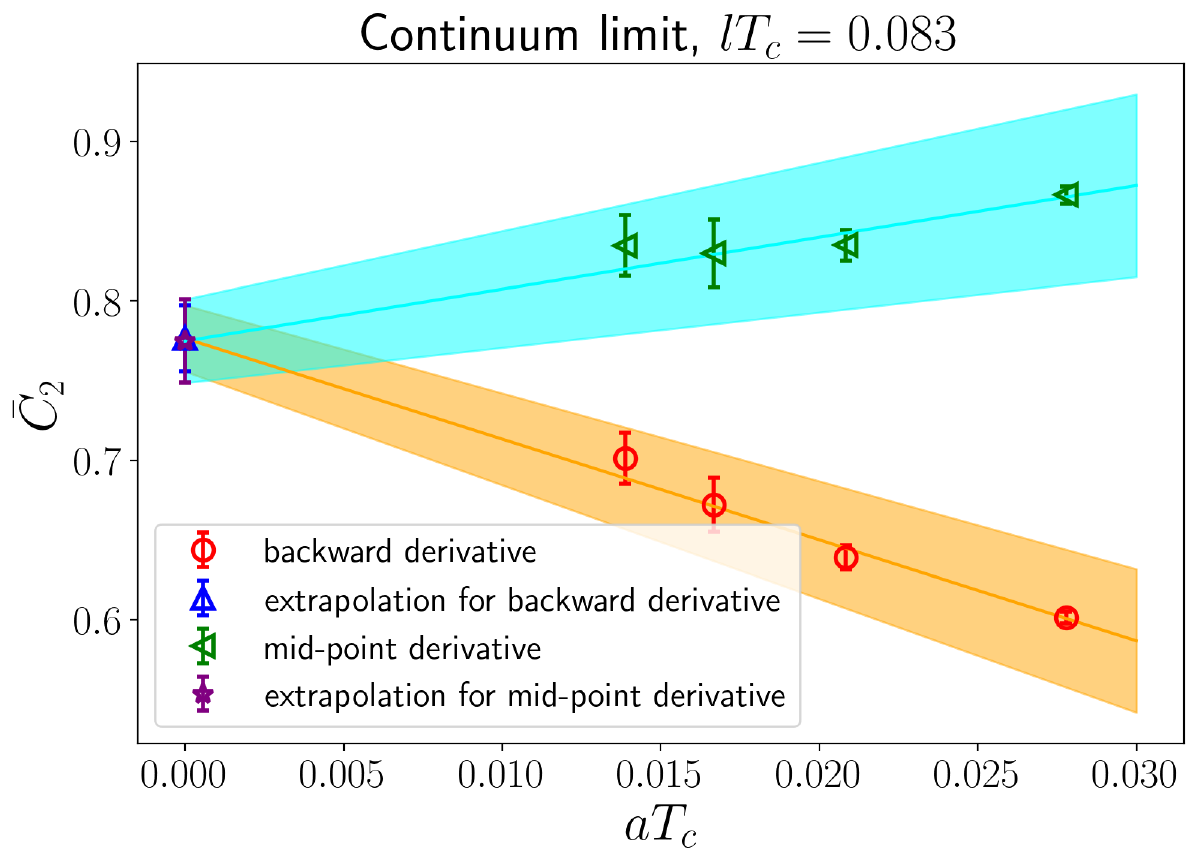}
\end{subfigure}
\begin{subfigure}{.47\textwidth}
\includegraphics[width=\textwidth]{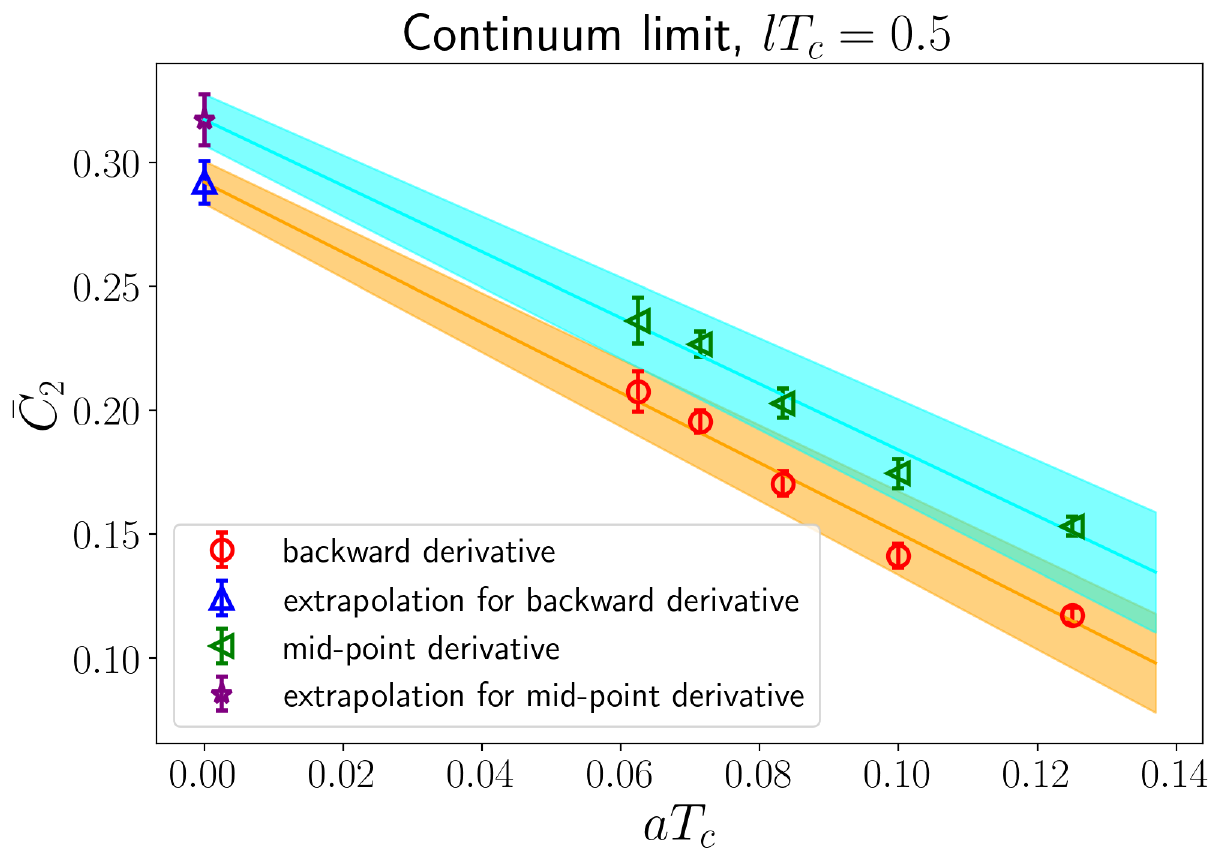}
\end{subfigure}
\begin{subfigure}{.47\textwidth}
\includegraphics[width=\textwidth]{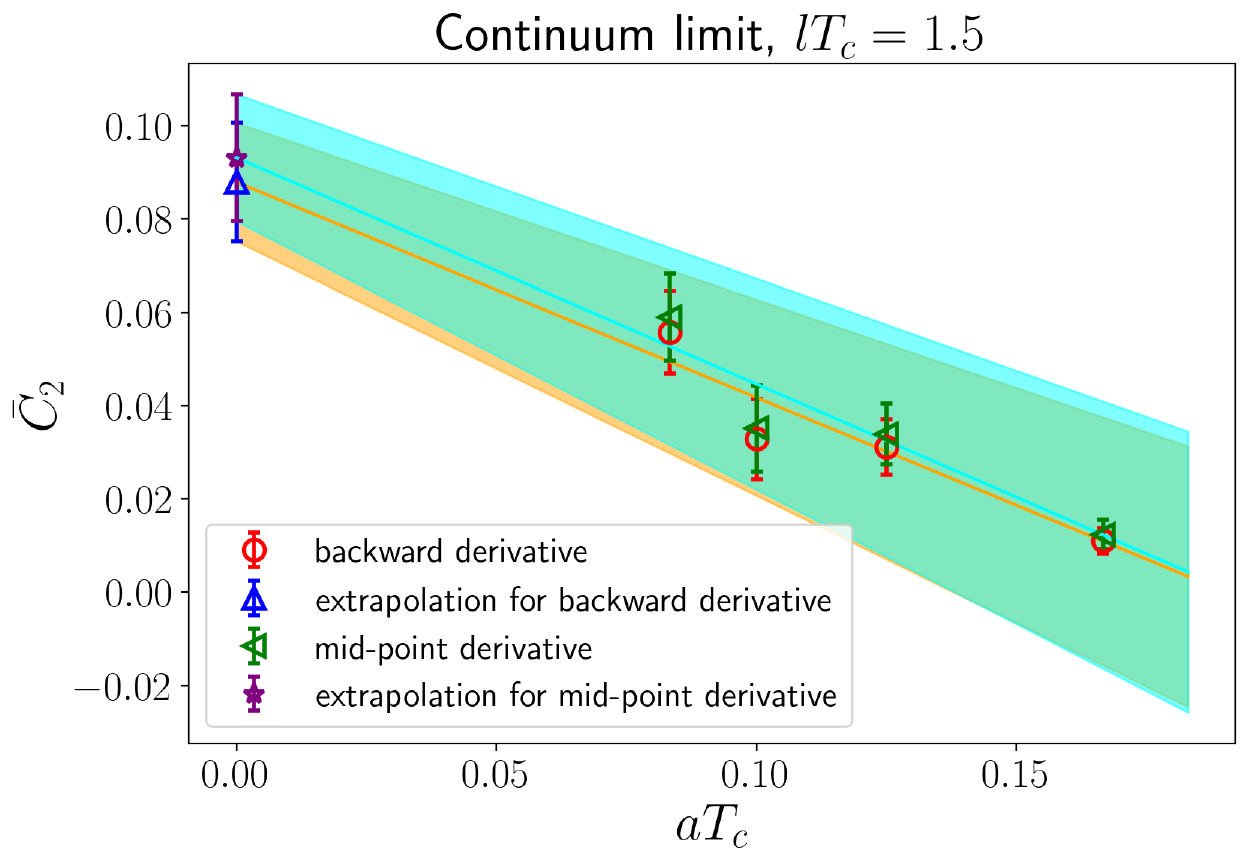}
\end{subfigure}
\caption{Different continuum limits.}
\label{fig:continuum_limit}
\end{figure}
The continuum extrapolation was performed in two different ways, as shown in figure~\ref{fig:continuum_limit}. The first one consists in approximating the derivative with the backward difference,
\begin{align}
\frac{1}{a}\ln\frac{Z_2(l)}{Z_2(l+a)} \simeq \eval{\frac{\partial S_2}{\partial l}}_{l}.
\end{align}
This approximation leads to discretization errors that are linear in the lattice spacing. The second method is to use the mid-point approximation,
\begin{align}
\frac{1}{a}\ln\frac{Z_2(l)}{Z_2(l+a)} \simeq \eval{\frac{\partial S_2}{\partial l}}_{l+a/2},
\end{align}
which, in principle, has $\order{a^2}$ discretization errors. However, with this approach the different points we use for the continuum extrapolations are evaluated at $(l+a/2)\mg$, which tends to $l\mg$ as $a$ goes to zero but introducing further discretization errors which are linear in $a$. We took the absolute difference between the continuum extrapolations carried out with these two methods as an estimate of (a source of) the systematic error on our results. The final result reported is the value of the mid-point approximation with an uncertainty obtained by summing in quadrature the statistical and the systematic errors. 

\begin{table}[t]
\begin{center}
\begin{small}
\begin{tabular}{|c|c|}
\hline
$N_{\tau, c}$ & $\beta$ \\
\hline\hline
$6$ & $0.228818(4)$ \\
$8$ & $0.226102(5)$ \\
$10$ & $0.224743(5)$ \\
$12$ & $0.223951(3)$ \\
$14$ & $0.223442(4)$ \\
$16$ & $0.223101(2)$ \\
$18$ & $0.2228492(15)$ \\
\hline
\end{tabular}
\begin{tabular}{|c|c|}
\hline
$N_{\tau, c}$ & $\beta$ \\
\hline\hline
$20$ & $0.2226632(13)$ \\
$24$ & $0.2224077(9)$ \\
$25$ & $0.2223601(8)$ \\
$28$ & $0.2222431(7)$ \\
$30$ & $0.2221817(7)$ \\
$36$ & $0.2220486(5)$ \\
$40$ & $0.2219876(4)$ \\
\hline
\end{tabular}
\begin{tabular}{|c|c|}
\hline
$N_{\tau, c}$ & $\beta$ \\
\hline\hline
$45$ & $0.2219306(3)$ \\
$48$ & $0.2219037(3)$ \\
$50$ & $0.2218880(3)$ \\
$60$ & $0.2218292(2)$ \\
$72$ & $0.22178524(16)$ \\
$75$ & $0.22177703(15)$ \\
$90$ & $0.22174622(12)$ \\
\hline
\end{tabular}
\end{small}
\end{center}
\caption{Values of $\beta$ associated to each $N_{\tau,\mbox{\tiny{c}}}$.}
\label{tab:critical_values}
\end{table}

\begin{table}[t]
\begin{center}
\begin{small}
\begin{tabular}{|c|c|c|}
\hline
$l\Tc$ & $N_{\tau,c}$ & $N_s$ \\
\hline\hline
$0.066$ & $[45,60,75,90]$ & $[100;160],[100;160],[90;180],[130;210]$ \\
\hline
$0.083$ & $[36,48,60,72]$ & $[80;155],[90;165],[100;160],[105;165]$\\
\hline
$0.1$ & $[30,40,50]$ & $[70;150],[70;160],[80;130]$\\
\hline
$0.166$ & $[18,24,30]$ & $[50;130],[100;140],[80;140]$\\
\hline
$0.2$ & $[15,20,25]$ & $[40;90],[50;90],[50;110]$ \\
\hline
$0.25$ & $[12,20,24,28]$ & $[50;80],[50;120],[100;140],[50;130]$ \\
\hline
$0.333$ & $[15,18,21,24]$ & $[40;65],[45;70],[50;70],[50;90]$ \\
\hline
$0.375$ & $[8,16,24]$ & $[25;50],[45;78],[40;120]$ \\
\hline
$0.5$ & $[8,10,12,14,16]$ & $[24;48],[28;56],[40;68],[44;72],[52;72]$ \\
\hline
$0.625$ & $[8,12^\star,16]$ & $[28;52],[36;68],[40;74]$ \\
\hline
$0.75$ & $[8,12,16]$ & $[28;52],[44;68],[54;70]$ \\
\hline
$0.875$ & $[8,12^\star,16]$ & $[54],[76],[80]$\\
\hline
$1$ & $[8,10,12,16]$ & $[36;54],[65],[68],[40;80]$ \\
\hline
$1.125$ & $[6^\star,8,10^\star]$ & $[70],[70],[35;80]$ \\
\hline
$1.25$ & $[8,12,16]$ & $[58],[72],[100]$ \\
\hline
$1.375$ & $[6^\star,8,10^\star]$ & $[70],[70],[80]$ \\
\hline
$1.5$ & $[6,8,10,12]$ & $[70],[70],[80],[80]$ \\
\hline
\end{tabular}
\end{small}
\end{center}
\caption{Details of our simulations. First column: length of the slab in units of the inverse of the critical temperature. Second column: different critical temporal extents (i.e., lattice spacings) used for the continuum extrapolation; for the values marked with a star, we calculated the entropic c-function for two different lengths of the cut and interpolated linearly to the corresponding $l\Tc$. Third column: ranges of spatial extents used for the thermodynamic extrapolation.}
\label{tab:thermodynamic_limit}
\end{table}

\bibliography{paper}

\end{document}